\documentclass[preprint,12pt]{elsarticle} 

\usepackage[utf8]{inputenc}

\usepackage{ulem}

\usepackage{amssymb}
\usepackage{tikz}
\usepackage{pgfplots}
\usepackage{epstopdf}
\pgfplotsset{compat=1.14}
\usetikzlibrary{backgrounds,arrows,automata,shapes,positioning,calc,through,spy}
\pgfdeclarelayer{background}
\pgfdeclarelayer{foreground}
\pgfsetlayers{background,main,foreground}
\usepgfplotslibrary{colormaps}

\tikzset{
  imglabel/.style={
    rectangle,
    inner sep=2pt,
    rounded corners=.1em,
    text=black,
    minimum height=1em,
    text centered,
    fill=white,
    fill opacity=1.0,
    text opacity=1,
    anchor=south west,
    draw=black,
  },
  imgletter/.style={
    rectangle,
    inner sep=2pt,
    text=black,
    minimum height=1em,
    text centered,
    fill=white,
    fill opacity=0.7,
    text opacity=1,
    anchor=south west,
  },
  double arrow/.style args={#1 colored by #2 and #3}{
    -stealth,line width=#1,#2, 
    postaction={draw,-stealth,#3,line width=(#1)/3,
    shorten <=(#1)/3,shorten >=2*(#1)/3}, 
  }
}

\usepackage[export]{adjustbox}

\journal{Acta Astronautica}
\usepackage{subcaption}
\usepackage{siunitx}
\DeclareSIUnit \parsec {pc}
\DeclareSIUnit \electronvolt {eV}
\DeclareSIUnit \pixel {px}
\DeclareSIUnit \arcmin {arcmin}
\DeclareSIUnit \erg {erg}

\newcommand{\reffig}[1]{Figure~\ref{#1}}
\newcommand{\reftab}[1]{Table~\ref{#1}}

\usepackage[]{hyperref}

\usepackage[nolist,nohyperlinks]{acronym}
\acrodef{REX}[REX]{Rocket Experiment}
\acrodef{WRXR}[WRXR]{Water Recovery X-ray Rocket}
\acrodef{FOV}[FOV]{Field of View}
\acrodef{LE}[LE]{Lobster-Eye}
\acrodef{TRL}[TRL]{Technology Readiness Level}
\acrodef{SNR}[SNR]{Supernova Remnant}
\acrodef{MFO}[MFO]{Multi-Foil Optics}
\acrodef{CMOS}[CMOS]{Complementary Metal–Oxide–Semiconductor}
\acrodef{PSU}[PSU]{Pennsylvania State University}
\acrodef{IR}[IR]{infrared}
\acrodef{IMU}[IMU]{Inertial Measurement Unit}
\acrodef{ROS}[ROS]{Robot Operating System}
\acrodef{CTU}[CTU]{Czech Technical University in Prague}
\acrodef{RITE}[RITE]{Rigaku Innovative Technologies}
\acrodef{PSF}[PSF]{Point Spread Function}
\acrodef{ASIC}[ASIC]{Application-Specific Integrated Circuit}
\acrodef{ADC}[ADC]{Analog-to-Digital Converter}
\acrodef{ADCS}[ADCS]{Attitude Determination and Control System}
\acrodef{MEMS}[MEMS]{Micro-electro-mechanical System}
\acrodef{UWB}[UWB]{University of West Bohemia}
\acrodef{1D}[1D]{one-dimensional}
\acrodef{2D}[2D]{two-dimensional}

\begin{document}


\begin{frontmatter}



  \title{REX: X-ray experiment on~the~Water Recovery Rocket}


  \author[ctu]{Martin Urban\corref{cor1}} \ead{martin-urban@fel.cvut.cz}
  \author[ctu]{Ondřej Nentvich} \ead{ondrej.nentvich@fel.cvut.cz}
  \author[ctu]{Tomáš Báča} \ead{tomas.baca@fel.cvut.cz}
  \author[zcu]{Ivo Veřtát} \ead{ivertat@fel.zcu.cz}
  \author[rite]{Veronika Maršíková} \ead{veronika.marsikova@rigaku.com}
  \author[rite,adva]{Daniela Doubravová} \ead{daniela.doubravova@advacam.com}
  \author[vzlu]{Vladimír Dániel} \ead{daniel@vzlu.cz}
  \author[ctu]{Adolf Inneman}
  \author[ctu,rite]{Ladislav Pína}
  \author[penn]{Randall L. McEntaffer} \ead{rlm90@psu.edu}
  \author[penn]{Ted B. Schultz}
  \author[penn]{Drew M. Miles}
  \author[penn]{James H. Tutt}
  
  \address[ctu]{Czech Technical University in Prague, Czech Republic}
  \address[zcu]{University of~West Bohemia, Pilsen, Czech Republic}
  \address[rite]{Rigaku Innovative Technologies Europe, s.r.o., Prague, Czech Republic}
  \address[adva]{Advacam, s.r.o., Prague, Czech Republic}
  \address[vzlu]{Czech Aerospace Research Centre a.s., Prague, Czech Republic}
  \address[penn]{Pennsylvania State University, State College, PA, USA}

  \cortext[cor1]{Corresponding author}

  \begin{abstract}
    This paper presents \ac{REX} that was part of a dual-payload rocket campaign for NASA's sounding rocket Black Brant~IX with water recovery technology.
    This mission was a suborbital sounding rocket flight that was launched and recovered on April~4,~2018 and targeted the~Vela supernova remnant.
    The~purpose of~\ac{REX} was to classify the~\acl{TRL} of onboard devices designed for space applications.
    The devices were two wide-field X-ray telescopes consisting of a combination of \ac{LE} optics with an uncooled Timepix detector (256\,$\times$\,256\,\si{\pixel} @ \SI{55}{\micro\meter}), and additional sensors.
    The first telescope uses a~\acl{2D} combination of \ac{LE} modules with a focal length of~\SI{1}{\meter} and a~\ac{FOV} of~1.0\,$\times$\,1.2\,\si{\degree} and operates in the energy range of~3\,--\,60\,\si{\kilo\electronvolt}.
    The~second telescope was a~\acl{1D} \ac{LE} with a~focal length of~\SI{250}{\milli\meter} and a~\ac{FOV} of~2.7\,$\times$\,8.0\,\si{\degree} for the~energy range 3\,--\,40\,\si{\kilo\electronvolt}.
    The~X-ray telescopes were supplemented by a~camera in~the~visible spectrum with 1,280\,$\times$\,1,024\,\si{\pixel} resolution, which was used to obtain images of~the~observed sources and to verify the~resulting pointing of~the~rocket carrier.
    Other devices also include infrared array sensors and inertial measurement units tested for future small satellite missions.
    The~data handler and communication system were built using~the~Robot Operating System, and both the~system and the~electronics were deployed and operated in-flight.
    The hardware was successfully recovered after the launch and the data were extracted.
  \end{abstract}


  


\end{frontmatter}




\section{Introduction}\label{sec:Intro}

\acresetall


Sounding rockets, a type of Suborbital rockets, provide a~low-cost observing platform for carrying instruments for scientific and educational purposes, as well as for rapid technology validation.
This is very advantageous for testing the~\ac{TRL} of space instrumentation without the need for expensive boosters and advanced telemetry systems.
Therefore, the~price of~the launch is lower than for satellite missions.
The~disadvantage of~using suborbital rockets is in the~limited experiment time, which is usually 5\,--\,20\,\si{\minute}.
Nevertheless, they are still useful for verification of~technologies in~space which do not require a~long experimental time.
A~major consideration for the~\ac{TRL} of~devices used in~space is not only the~capability of~working in~vacuum, extreme temperatures, temperature changes, and radiation, but also the~ability to survive the~violent conditions during launch and booster separation \citep{Christe2016}.

Aside from \ac{TRL} verification, sounding rockets can also be used for scientific purposes.
One example of~a~scientific mission was the~CHAMPS mission launched in~October, 2011.
Two rockets were used to detect and measure the~density and size distribution of~meteoritic smoke particles in~the~upper atmosphere~\citep{Robertson2014}.
Another mission was focused on~the~examination of~the~Sun's atmosphere and was launched in~2009 by NASA's sounding rocket.
Its goal was the~research of~the~helium amount and how the~solar wind originated and accelerated~\citep{Moses2020}.
The~last mentioned example is from the \emph{RockOn!} series of sounding rocket missions carrying experiments developed by the~students during NASA supported workshops\footnote{\href{https://www.nasa.gov/wallops/2019/feature/nasa-launches-student-experiments-to-space-on-a-suborbital-rocket}{https://www.nasa.gov/wallops/2019/feature/nasa-launches-student-experiments-to-space-on-a-suborbital-rocket}}\footnote{\url{https://spacegrant.colorado.edu/national-programs/rockon-home}}.
Rockets were launched from Wallops Flight Facility in~Virginia regularly since~2008~\citep{Dirienzo2017,Revitalizing2010}.

\begin{figure}
  \centering 
  \begin{subfigure}[b]{0.32\textwidth}
    \includegraphics[width=\textwidth]{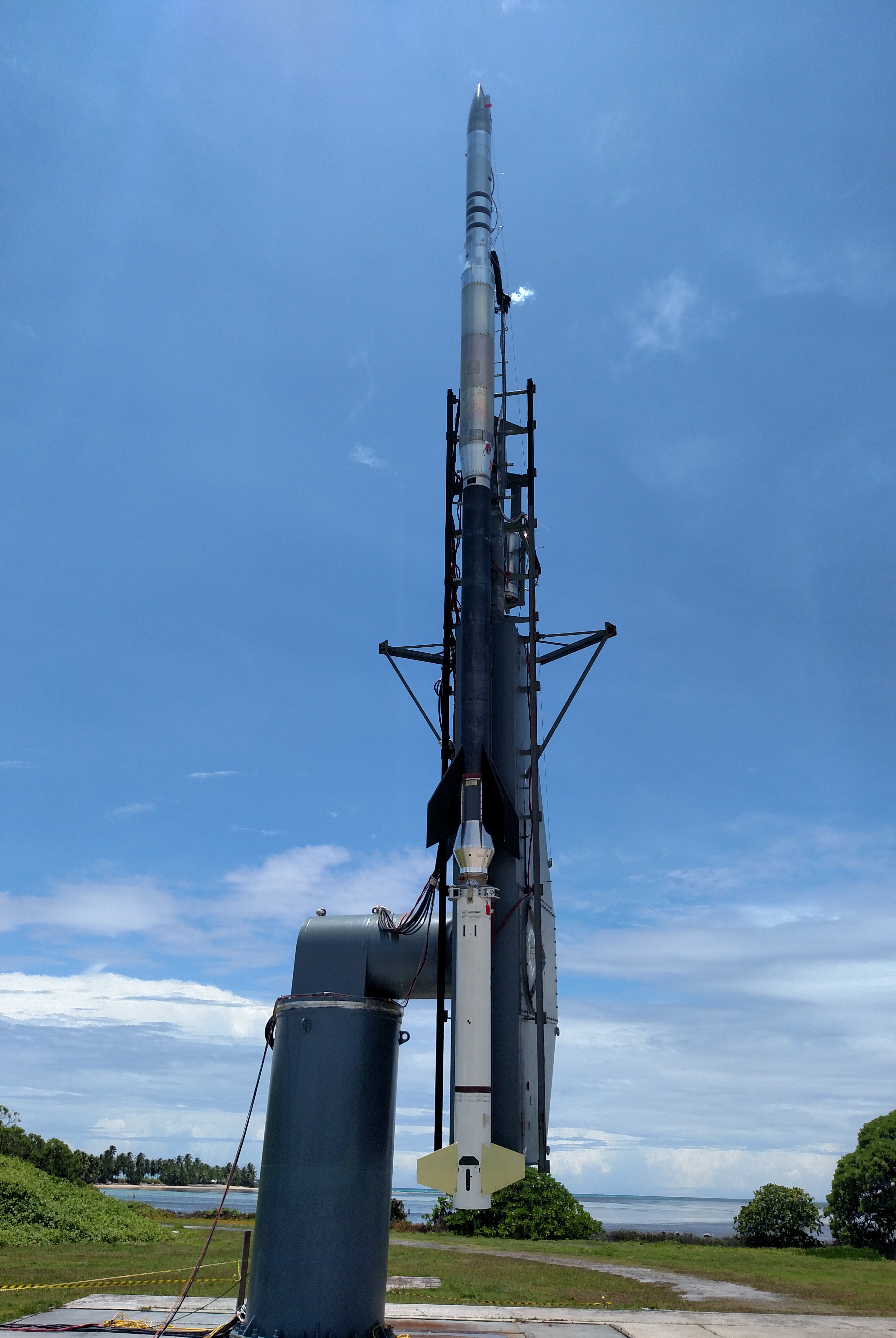}
    \caption{At the launch site.}
  \end{subfigure}
  \hfill
  \begin{subfigure}[b]{0.32\textwidth}
    \includegraphics[width=\textwidth]{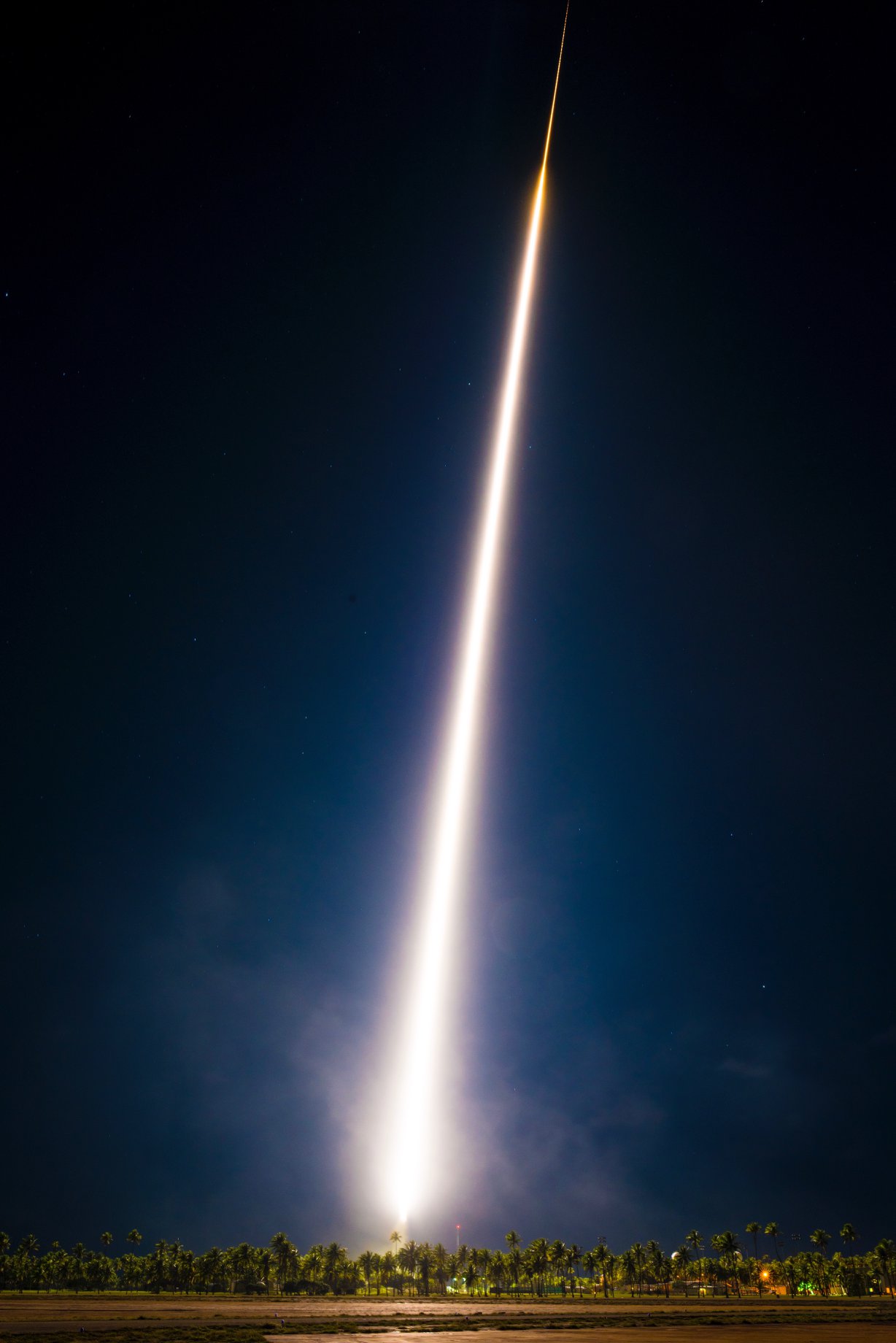}
    \caption{During the launch.}
  \end{subfigure}
  \hfill%
  \centering 
  \begin{subfigure}[b]{0.32\textwidth}
    \includegraphics[width=\textwidth]{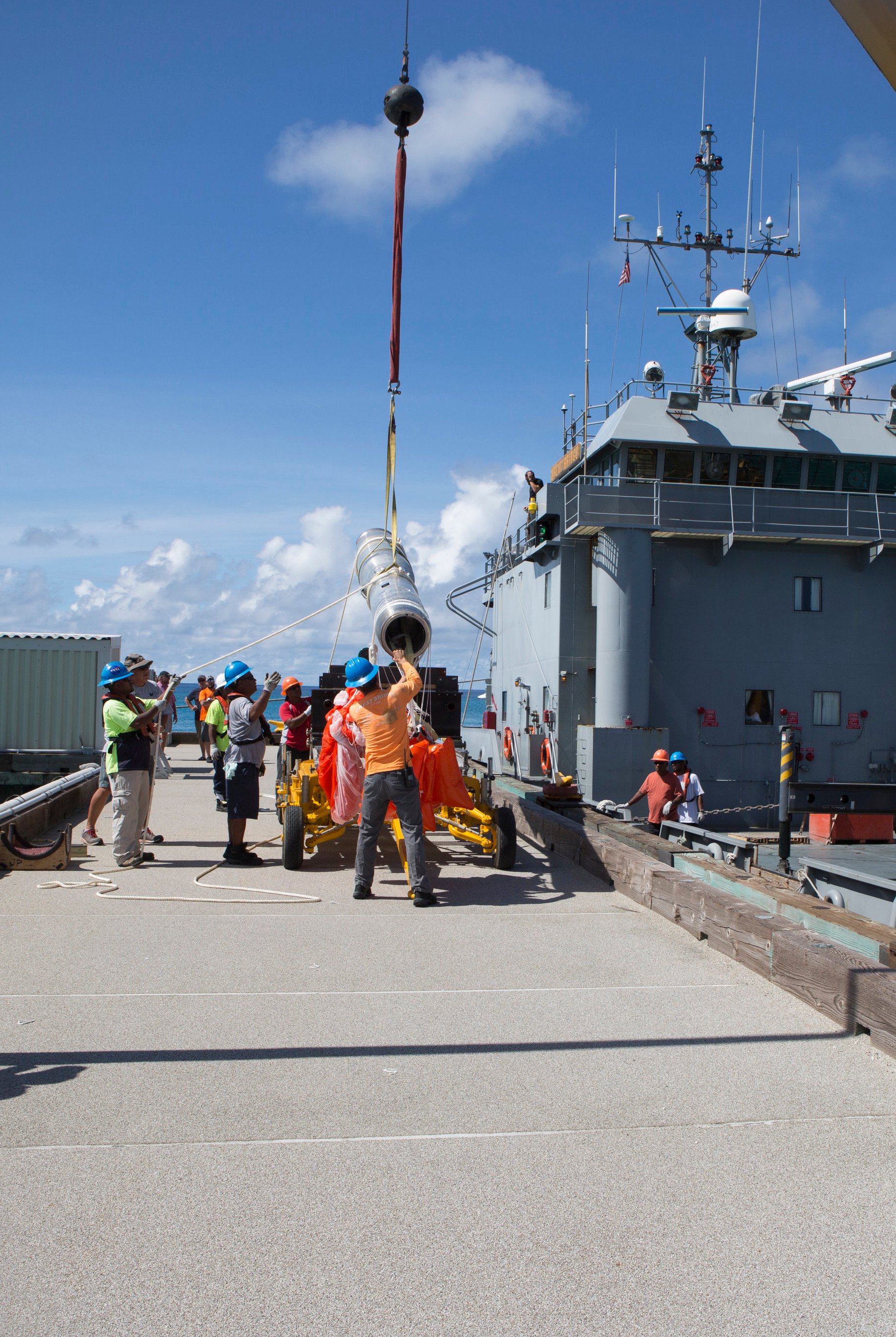}
    \caption{Recovery at sea.}
  \end{subfigure}
  \caption{Rocket carrier and its launch from Kwajalein Atoll in~the~Marshall Islands.}
  \label{fig:Rocket}
\end{figure}

The suborbital mission described in~this manuscript was a~two-payload campaign carried by the~NASA's sounding rocket with water recovery technology for science payloads which was launched from the~Kwajalein Atoll in the Marshall Islands on April~4,~2018. 
Figure~\ref{fig:Rocket} shows photos from the launch date.


The observation of~the~Vela \ac{SNR} was the~primary scientific astrophysical target of~this sub-orbital experiment.
The~Vela \ac{SNR} is a~shell-type remnant located in~the~southern constellation Vela at~a~distance of~approx. \SI{250}{\parsec} \citep{Cha1999}.
This source is Type II and it exploded approximately 11,000\,--\,12,300 years ago.
The centre of the Vela \ac{SNR} is formed by a~neutron star with a~strong magnetic field, with a~mass close to our Sun and a~diameter of~approx.~\SI{20}{\kilo\meter}~\citep{Daniel2019} (the apparent diameter of~$\approx$\,\SI{8}{\degree} \citep{Miles2017}).
Due to the apparent size of the object and high surface brightness, this \ac{SNR} is a~frequent source of~observation and testing of~equipment for both sounding rockets \citep{Seward2010,Gorenstein1974,Burkert1982} and space telescopes \citep{Kahn1985,Bamba2005}.
In addition to scientific observation, the~mission aims at an important technological goal, namely the~suborbital flight verification of~scientific payloads such as the~\ac{MFO} type of~\ac{LE} with a~Timepix pixel detector \citep{Llopart2007}, and X-ray reflection gratings with a~hybrid \ac{CMOS} detector.

The sounding rocket was prepared by the~\ac{PSU} together with the~main measuring instrument --- a~soft X-ray grating spectrometer (bandpass 0.25\,--\,0.75\,\si{\kilo\electronvolt}) named the~\ac{WRXR} payload.
The~spectrometer consisted of~a~mechanical collimator, X-ray reflection gratings, grazing incidence mirrors, and a~hybrid \ac{CMOS} detector.
In-depth description of~the~mission goals and the \ac{WRXR} payload can be found in~\citep{Miles2017,Miles2019,Tutt2019,Wages2019}.


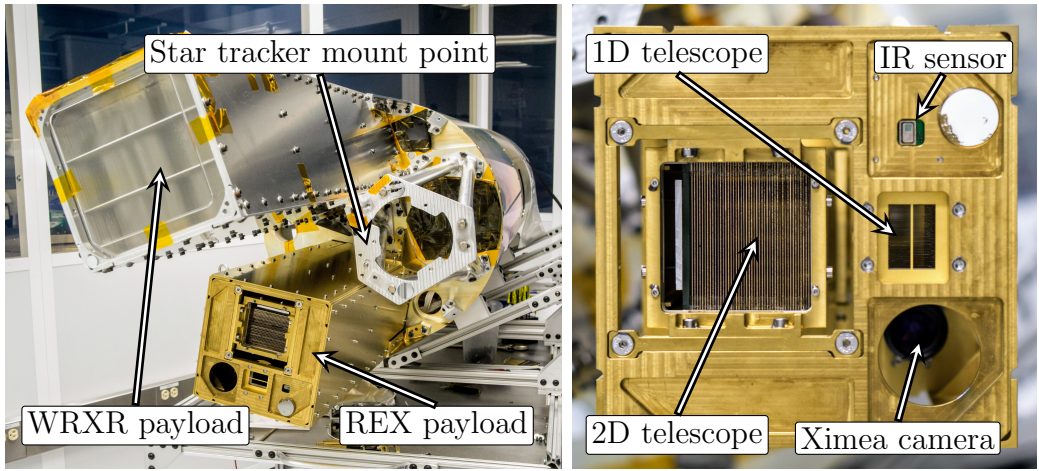
\begin{figure*}
  \centering 
  \begin{subfigure}{0.54\textwidth}
    \begin{tikzpicture}
      \node[anchor=south west,inner sep=0] (a) at (0,0) { \adjincludegraphics[width=1.0\textwidth,trim={{0.15\width} {0.20\height} {0.25\width} {0.05\height}}, clip]{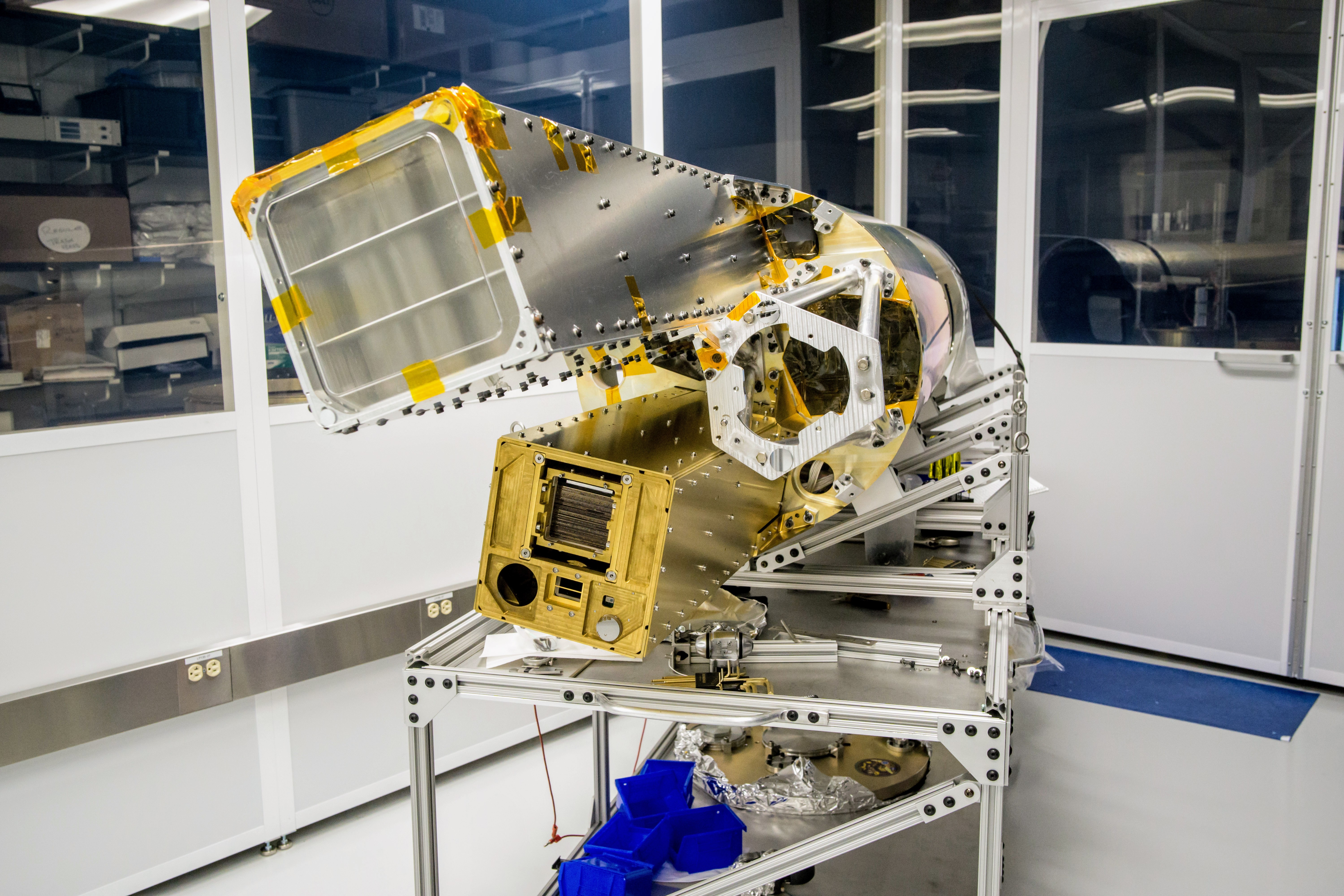}};
      \begin{scope}[x={(a.south east)},y={(a.north west)}]
        \node[imglabel,text=black] (wrxr_label) at (0.03, 0.05) {\normalsize WRXR payload};
        \draw[double arrow=3pt colored by black and white,rounded corners] (wrxr_label.north) -- (0.28, 0.65);

        \node[imglabel,text=black] (rex_label) at (0.60, 0.05) {\normalsize REX payload};
        \draw[double arrow=3pt colored by black and white,rounded corners] (rex_label.north) -- (0.55, 0.25);

        \node[imglabel,text=black] (st_label) at (0.25, 0.85) {\normalsize Star tracker mount point};
        \draw[double arrow=3pt colored by black and white,rounded corners] (st_label.south) -- (0.65, 0.48);
      \end{scope}
    \end{tikzpicture}
    \caption{The sensor and optic compartment of the rocket payload.}
    \label{fig:rex_and_wrxr}
  \end{subfigure}
  \begin{subfigure}{0.45\textwidth}
    \begin{tikzpicture}
      \node[anchor=south west,inner sep=0] (a) at (0,0) { \adjincludegraphics[width=1.0\textwidth]{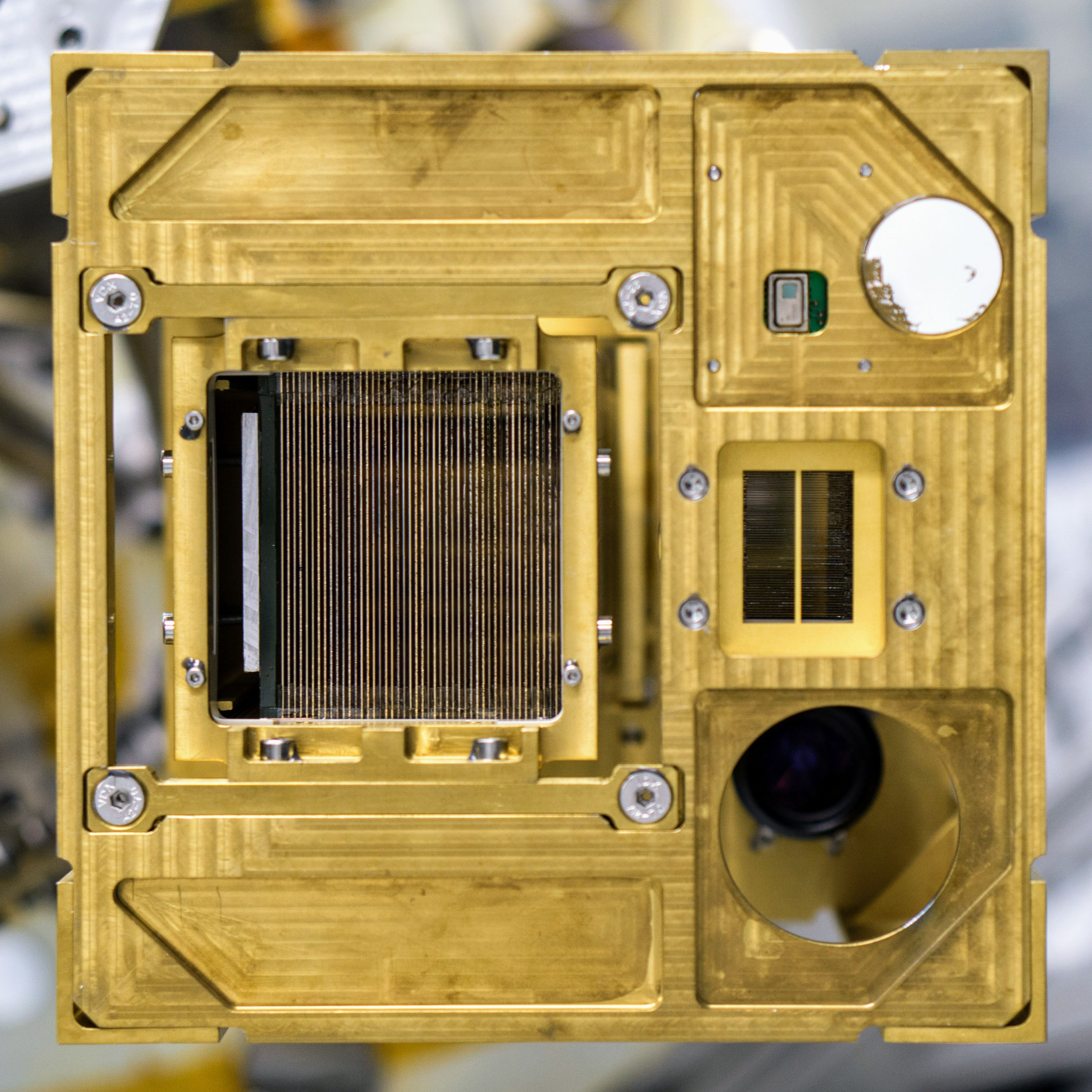}};
      \begin{scope}[x={(a.south east)},y={(a.north west)}]
        \node[imglabel,text=black] (2d_label) at (0.03, 0.03) {\normalsize 2D telescope};
        \draw[double arrow=3pt colored by black and white,rounded corners] (2d_label.north) -- (0.40, 0.50);

        \node[imglabel,text=black] (1d_label) at (0.03, 0.85) {\normalsize 1D telescope};
        \draw[double arrow=3pt colored by black and white,rounded corners] (1d_label.south) -- (0.70, 0.50);

        \node[imglabel,text=black] (ximea_label) at (0.48, 0.03) {\normalsize Ximea camera};
        \draw[double arrow=3pt colored by black and white,rounded corners] (ximea_label.north) -- (0.75, 0.28);

        \node[imglabel,text=black] (ir_label) at (0.65, 0.85) {\normalsize IR sensor};
        \draw[double arrow=3pt colored by black and white,rounded corners] (ir_label.south) -- (0.74, 0.75);
      \end{scope}
    \end{tikzpicture}
    \caption{Apertures at the front of the \acs{REX} instrument.}
    \label{fig:rex_apertures}
  \end{subfigure}
  \caption{The final arrangement of~the~rocket vacuum segment with \acs{REX} payload (\acs{1D} and \acs{2D} \acs{LE} telescopes) and \acs{WRXR} payload (soft X-ray spectroscope).}
  \label{fig:REXinLab}
\end{figure*}


The second payload --- the~\ac{REX} --- was prepared by the~Czech team and contained two sets of~\ac{MFO} in~combination with the~Timepix detector, which formed two separate X-ray telescopes.
The~first of the \ac{REX} telescopes consists of~two \ac{LE} modules in~the~Schmid's arrangement \citep{Schmidt1975,Angel1979} which covered a~\ac{FOV} of~1.0\,$\times$\,1.22\,deg in~the~energy range of~3\,--\,60\,\si{\kilo\electronvolt}.
The second telescope has a single \ac{1D} \ac{LE} module \citep{Hudec2010,Pina2018} with a~field of~view of~\,2.75\,$\times$\,8.0\,\si{\degree} for the~energy range 3\,--\,40\,keV. 
Moreover, the payload was equipped with additional sensors: \emph{Ximea} camera for visible spectrum with the~resolution of~1,280\,$\times$\,1,024\,\si{\pixel}, an~\ac{IR} sensor array and an \ac{IMU}.
A photo of the payload is shown in \reffig{fig:REXinLab} with the~\ac{WRXR} on top and the \ac{REX} on the bottom.



\section{REX instrument description}\label{sec:REX}

The \ac{REX} payload consisted of~two main parts: the~optics modules including sensors, and a~hermetically sealed electronics box.
An~airtight bulkhead separated both parts and they were electrically connected via a~vacuum feedthrough.
The payload was also connected to a rocket interface panel via another vacuum feedthrough.

\begin{figure}[b!]
  \centering 
  \includegraphics[width=.85\textwidth]{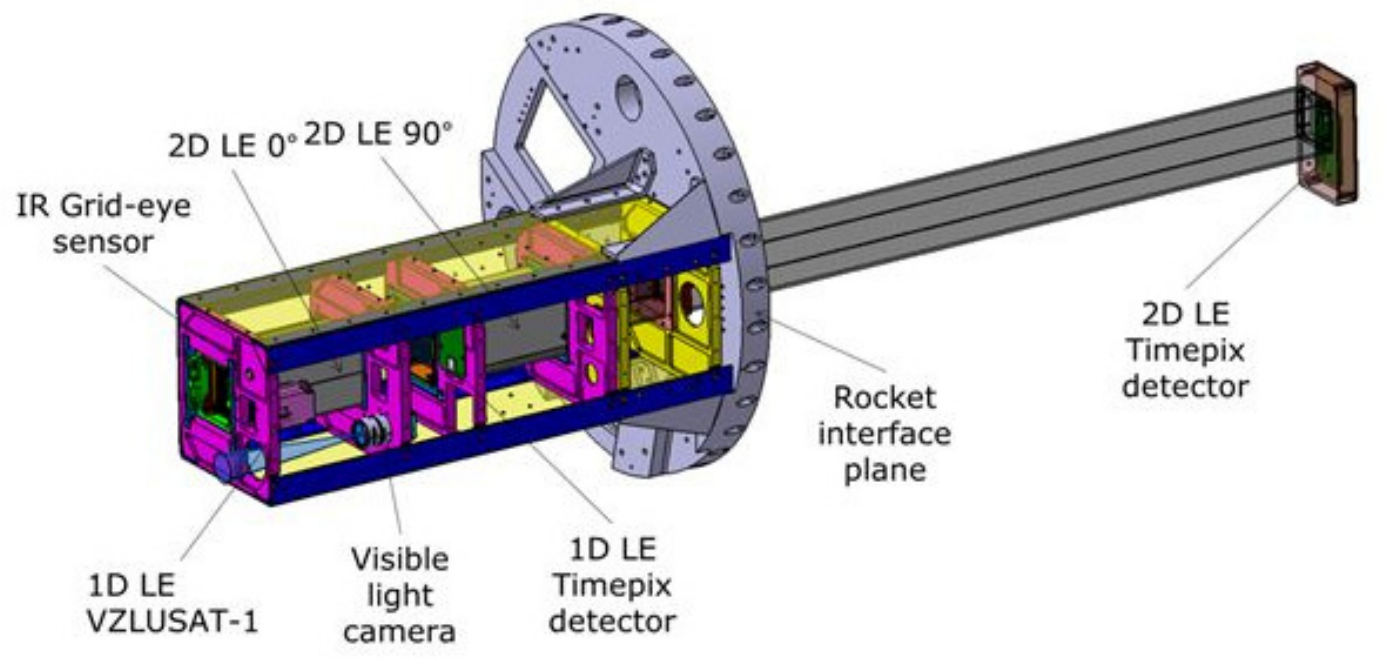}
  \caption{3D rendition of~the~vacuum part of~the~X-ray optical system \acs{REX}.}
  \label{fig:model}
\end{figure}

The optics and sensors section contained all the~sensors and detectors except for the~\ac{IMU}, which was placed in~the~hermetic box.
The~sensor section was designed as hermetically sealed in~order to hold an~artificial low-pressure atmosphere to prevent liquid condensation during launch.
The visible-light camera was used alongside the~X-ray telescopes as well as the \ac{IR} sensor array. 
A~carbon-fiber baffle covered the optical path between the~optic modules and Timepix detectors to avoid side rays.
A~schematic of the payload is shown in~\reffig{fig:model}.


The hermetic box contained NiCd batteries and switching power supply electronics with DC/DC converters for powering the~devices during the~entire duration of~the~experiment (approx. 2~hours).
Two on-board Odroid-XU4 ARM computers managed the sensors and miscellaneous electronics provided an~interface with the~rocket.
Each of~the~computers controlled a~single Timepix detector (one per telescope), and the~rest of~the~sensors were evenly distributed between the~computers.
Mission software using \ac{ROS}\footnote{\url{http://ros.org}} was developed to operate and control the~entire experiment.
\ac{ROS} is a~widely adopted open-source middleware for the~integration of~sensors and data processing algorithms for autonomous systems.
Simultaneously, an~operating interface for Timepix detectors in \ac{ROS} was developed under the name \emph{Rospix}\footnote{\url{http://github.com/rospix/rospix}} \citep{Baca2018}.


\subsection{X-ray telescope based on \acl{LE} optics and Timepix detector}\label{subsec:LEandTPX}

The \ac{REX} experiment consisted of~two \ac{LE} telescopes designed by \ac{CTU}, manufactured by \ac{RITE}, a~\ac{2D} in Schmidt's arrangement and a~\ac{1D} \acl{LE} optics.
Table~\ref{tab:LE_parameters} presents all the~parameters of~the~optics: the~coating of~the~mirror layers, the~focal length or \ac{FOV} and the~physical position of~the~module in~the~experiment.
The~\ac{1D} \ac{LE} optics has the~same parameters as an identical copy used in~our previous successful mission on~the~nanosatellite VZLUSAT-1 \citep{Daniel2019,Urban2017}.

\begin{figure}[h]
  \centering
  \includegraphics[width=0.45\textwidth]{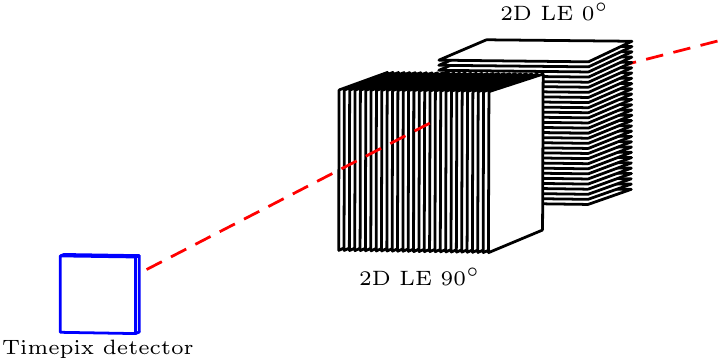}
  \caption{An illustration of the arrangement of the \acs{2D} optical system. Two \acs{1D} modules were combined in series with \SI{90}{\degree} mutual rotation along their optical axes.}
  \label{fig:2d_system_illustration}
\end{figure}

The position of~both \ac{1D} and \ac{2D} telescopes are visualised in~\reffig{fig:model}.
The~\ac{2D} telescope contained two \ac{1D} optical modules placed in series.
The modules were mutually rotated by \SI{90}{\degree} along their optical axes to form a \ac{2D} optical system.
Figure~\ref{fig:2d_system_illustration} shows an illustration of the~\ac{2D} system configuration.
A~Timepix detector was placed \SI{1438}{\milli\meter} from the~front aperture in the focal plane of~the~\ac{2D} system.
The~\ac{1D} telescope was placed alongside the \ac{2D}, with another Timepix detector in the focal plane.
The focal distance of the \ac{1D} was \SI{280}{\milli\meter} and the system contains only one optical module and thus the~values for vertical dimension are void.


\begin{table*}[t]
  \caption{\label{tab:LE_parameters}Parameters of~the~1D and 2D \acl{LE} optics for rocket experiment.}
  \centering
  \begin{tabular}{@{}lcc}
    \hline
    & \textbf{1D}                                 & \textbf{2D}                                  \\
    \hline
    Module centre position (horizontal) & \SI{250}{\milli\meter}                      & \SI{1363}{\milli\meter}                     \\
    Module centre position (vertical)   & -                                           & \SI{1097}{\milli\meter}                     \\
    Focal length (horizontal)           & \SI{243}{\milli\meter}                      & \SI{1345}{\milli\meter}                     \\
    Focal length (vertical)             & -                                           & \SI{1079}{\milli\meter}                                   \\
    Field of~view (horizontal)          & \SI{2.75}{\degree}                          & \SI{1.04}{\degree}                           \\
    Field of~view (vertical)            & $\approx$ \SI{8}{\degree}                   & \SI{1.29}{\degree}                           \\
    (limited by housing)                &                                             &                                              \\
    Mirror thickness                    & \SI{0.145}{\milli\meter}                    & \SI{0.35}{\milli\meter}                      \\
    Mirror spacing                      & \SI{0.3}{\milli\meter}                      & \SI{0.75}{\milli\meter}                      \\
    Mirror dimension                    & 60\,$\times$\,25\,\si{\milli\meter\squared} & 150\,$\times$\,75\,\si{\milli\meter\squared} \\
    Number of~mirrors                   & 56                                          & 2\,$\times$\,47                              \\
    Reflective surface                  & Au                                          & Au                                           \\
    Angular resolution (horizontal)     & \SI{6.12}{\arcmin}                          & \SI{1.40}{\arcmin}                           \\
    Angular resolution (vertical)       & -                                           & \SI{1.59}{\arcmin}                           \\
    \hline
  \end{tabular}
\end{table*}



\subsubsection{Simulating the~optical systems}\label{subsubsec:Simulation}

\begin{figure}
  \centering 
  \begin{subfigure}[b]{0.48\textwidth}
    \includegraphics[width=\textwidth]{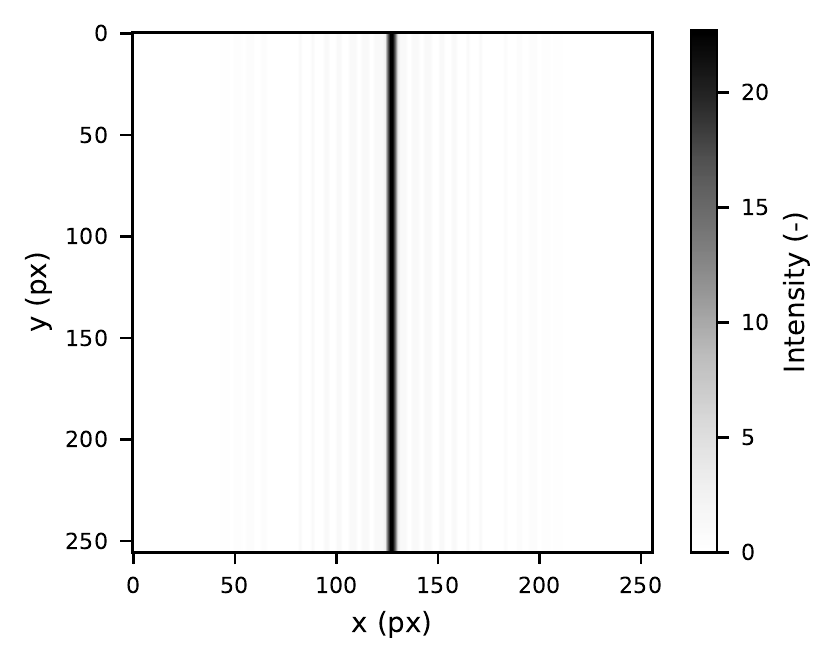}
    \caption{Simulating high flux.}
    \label{fig:simulated_1d_high_flux}
  \end{subfigure}
  \hfill
  \begin{subfigure}[b]{0.48\textwidth}
    \includegraphics[width=\textwidth]{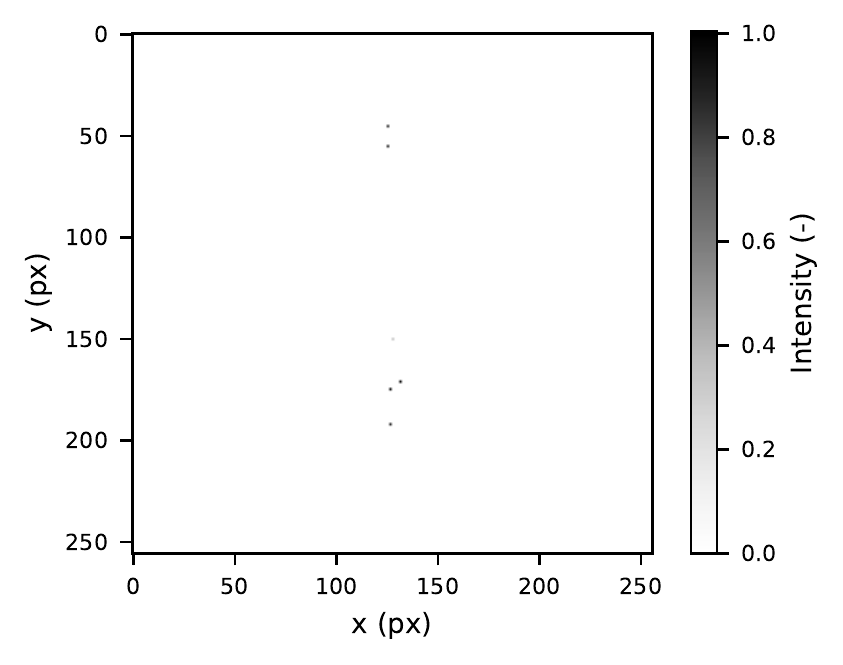}
    \caption{Simulating expected flux.}
    \label{fig:simulated_1d_expected_flux}
  \end{subfigure}
  \caption{(a) Simulated image with 100$\%$ mirror reflectivity for an~energy of~\SI{8}{\kilo\electronvolt} and a~uniform distribution of~the~photons on~the~input aperture and (b) calculated photons during the~observation time (\SI{282.5}{\second}).
  Both (a) and (b) images use 1D \acl{LE} optics with parameters listed in~\reftab{tab:LE_parameters}.}
  \label{fig:simulated_1d}
\end{figure}

\begin{figure}
  \centering 
  \begin{subfigure}[b]{0.48\textwidth}
    \includegraphics[width=\textwidth]{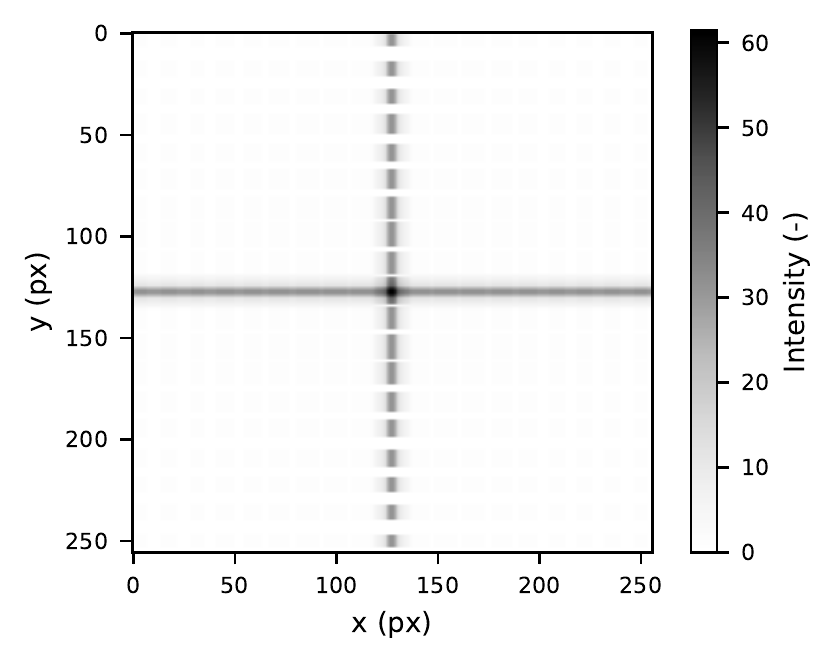}
    \caption{Simulating high flux.}
    \label{fig:simulated_2d_high_flux}
  \end{subfigure}
  \hfill
  \begin{subfigure}[b]{0.48\textwidth}
    \includegraphics[width=\textwidth]{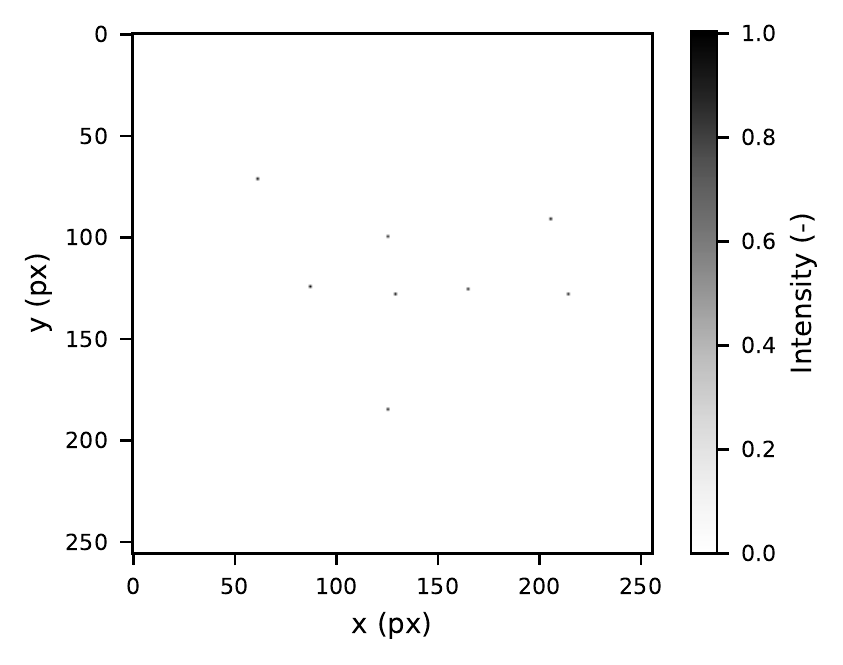}
    \caption{Simulating expected flux.}
    \label{fig:simulated_2d_expected_flux}
  \end{subfigure}
  \caption{(a) Simulated image with 100$\%$ mirror reflectivity for an~energy of~\SI{8}{\kilo\electronvolt} and a~uniform distribution of~the~photons on~the~input aperture and (b) calculated photons during the~observation time (\SI{282.5}{\second}).
  Both (a) and (b) images use 2D \acl{LE} optics with parameters listed in~\reftab{tab:LE_parameters}.}
  \label{fig:simulated_2d}
\end{figure}

The~Vela X-1 pulsar produces a~flux of~$3.0\cdot10^{-11}$ \si{\erg\per\centi\meter\squared\per\second} in~the~energy range of~4\,--\,25\,\si{\kilo\electronvolt} \citep{Willmore1992} to the~input aperture of~each optic.
For example, the~\ac{PSF} can be used with a~combination of~two \ac{1D} \acl{LE} optics for higher efficiency of~the~telescope, which leads to higher photon flux on the detector and thus in combination with a coded mask leads to a faster identification of~events in~the~sky as is proposed in~\citep{Nentvich2019}.
Figures \ref{fig:simulated_1d} and \ref{fig:simulated_2d} show simulated results of detector images where the~optics have 100$\%$\,reflectivity and the~detector 100$\%$ quantum efficiency.


\subsubsection{Timepix detector}\label{subsubsec:TPX}

\begin{figure}
  \centering 
  \begin{subfigure}[b]{0.55\textwidth}
    \includegraphics[width=\textwidth]{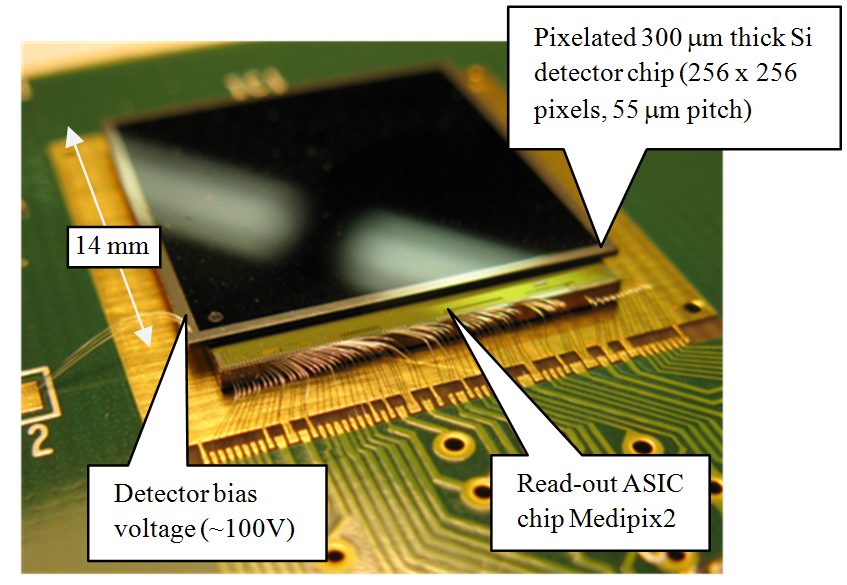}
    \caption{Timepix sensor with Si detector.}
    \label{fig:TPXchip}
  \end{subfigure}
  \hfill
  \begin{subfigure}[b]{0.375\textwidth}
    \begin{tikzpicture}
      \node[anchor=south west,inner sep=0] (a) at (0,0) { \includegraphics[width=1.0\textwidth]{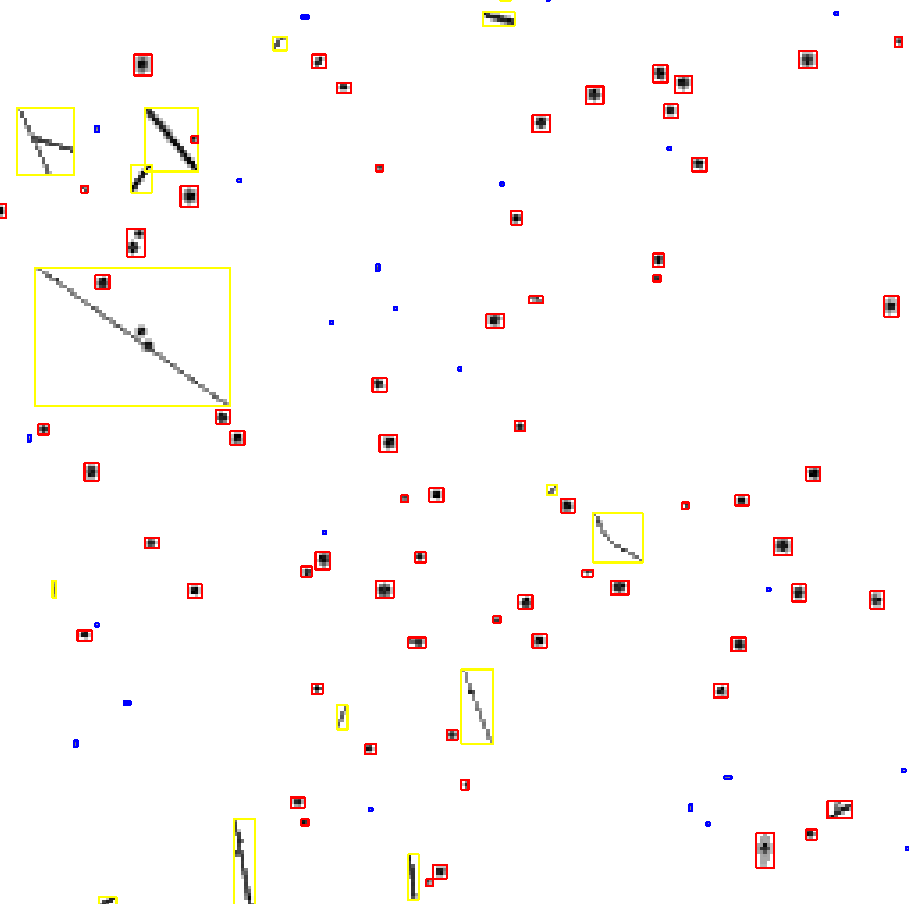}};
      \begin{scope}[x={(a.south east)},y={(a.north west)}]
        \draw (0.0, 0.0) rectangle (1.0, 1.0);
      \end{scope}
    \end{tikzpicture}
    \caption{An image of particle tracks.}
    \label{fig:TPXtrack}
  \end{subfigure}
  \hfill%
  \caption{Timepix (a) sensor consists of~a~\SI{300}{\micro\meter} Si detector and an~\acs{ASIC} read-out chip.
  The~sensor produces images with 256\,$\times$\,256\,\si{\pixel} resolution.
  (b) Ionizing particles leave characteristic marks (yellow --- electrons/muons, red --- ions, blue --- gamma), which can be processed by machine learning algorithms \cite{Baca2018} to extract the~particle type and its energy.}
  \label{fig:TPX}
\end{figure}

The Timepix detector \citep{Llopart2007} was used as an~X-ray detector for the~\ac{REX} experiment.
This hybrid silicon pixel detector developed at~CERN consisted of~a~high-density matrix of~256\,$\times$\,256 sensitive pixels with a~pixel pitch of~\SI{55}{\micro\meter}, and the~total sensitive area of~the~detector was 14.1\,$\times$\,14.1\,\si{\milli\meter\squared}.
The~Timepix sensor (see \reffig{fig:TPX}) consists of~two parts: the~first part is the~semiconductor detection layer (\SI{300}{\micro\meter} silicon in~this case used for the~energy range of~3\,--\,60\,\si{\kilo\electronvolt}), which is bump bonded to the~second part, an~\ac{ASIC} readout chip containing preamplifier, \ac{ADC}, and a~counter for each pixel.
Data from the~sensor can be read out as fast as 100 images per second (depending on~the~accompanying hardware interface \cite{vykydal2011usb, kraus2011fitpix, granja2018directional, trojanova2018evaluation}), which makes it ideal for particle tracking applications.
On the~other hand, long acquisition times on~the~order of~minutes are possible, thanks to the~sensor’s noise-less output.
The~matrix of~all 65,535 pixels can be equalized for a~given energy threshold which makes the~sensor filter out events with lower energy than the~threshold.
Thus it is possible to record even single-photon events while measuring their energy.

Timepix has already been extensively tested in~space \citep{Filgas2018}, e.g., onboard the~International Space Station \citep{Stoffle2015,Turecek2011}, the~Proba-V satellite \citep{Granja2016}, CubeSat VZLUSAT-1 \citep{Baca2016,Daniel2017,Urban2017}, TechDemoSat-1 \citep{Furnell2019,Hatfield2018}, and RISESAT \citep{Filgas2019}.
Two spare detectors from the~RISESAT project were used for the~\ac{REX} payload.


\subsection{Camera for the~visible spectrum}\label{subsec:Ximea}

\begin{figure}
  \centering 
  \hfill
  \begin{subfigure}[b]{0.3\textwidth}
    \includegraphics[width=\textwidth]{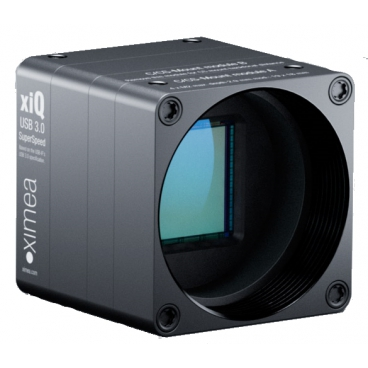}
    \caption{Ximea camera.}
  \end{subfigure}
  \hfill
  \begin{subfigure}[b]{0.3\textwidth}
    \includegraphics[width=\textwidth]{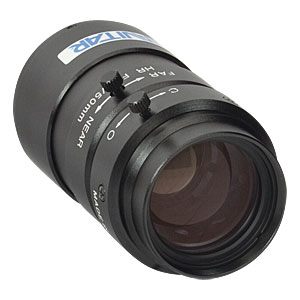}
    \caption{ThorLabs lens.}
  \end{subfigure}
  \hfill
  \mbox{}
  \caption{Small \ac{CMOS} camera module from Ximea with ThorLabs lens, used in~the~\ac{REX} experiment as star tracker.}
  \label{fig:Ximea} 
\end{figure}

The camera system for the visible spectrum was employed to verify the \ac{REX}’s target pointing in addition to the primary \ac{ADCS} of Black Brant~IX rocket.
Verification of~the~small, cheap, and uncooled industrial class \ac{CMOS} camera was the~secondary purpose of~this system, as it is considered for the~future CubeSat missions.
The~camera system was based on~the~Ximea MQ013CG-E2 module with a~small size (26\,$\times$\,26\,$\times$\,26\,\si{\milli\meter\cubed}) and low weight (\SI{26}{\gram}).
The~camera has a 1/1.8'' \ac{CMOS} sensor with \SI{1.3}{\mega\pixel} native resolution.
The~camera module was equipped with ThorLabs MVL50M23 lens, with a~focal length of~\SI{50}{\milli\meter}, offering a~field of~view approximately \SI{9.8}{\degree}.
Figure~\ref{fig:Ximea} shows the camera module and the used lens.
The~low power consumption and the~aluminum camera case attached to the~massive rocket mechanical structure mitigated the~possibility of~the~camera overheating in~vacuum.
An~absence of~plastic materials in~the~camera body avoided the~pollution of~the~telescopes by material outgassing.
The~USB3 interface and the~availability of~drivers for the~ROS made it easier to implement the~camera control software in~the~Odroid on-board computer.
The~camera outputs uncompressed RAW images which allow post-processing of~captured images after the~rocket's recovery.


\subsection{Additional sensors}

The \acl{IR} array sensor \emph{Grid-EYE} AMG88 (8\,$\times$\,8\,\si{\pixel} resolution) was added to the~\ac{REX} payload as a~technology demonstrator of Earth horizon detection.
It could be used as one of the inputs to \ac{ADCS} algorithms of a small CubeSat satellites.
Furthermore, we employed an off-the-shelf \ac{IMU}, the MPU6000.
It is a~miniature low cost 3-axis \ac{MEMS} accelerometer and gyroscope, widely used in~commercial electronics and robotics.
As this class of~electronics is not dedicated for space application, the~main reason for using it in~this mission was to test the~capability of~the~sensor to survive the~hard conditions of~the~rocket launch.
The~possibility of~its future usage for recognition of~flight phases and experiments triggering as well as future usage in~attitude determination and control systems of~small satellites will be examined.




\section{Experimental results}


\subsection{Launch event and payload recovery}

The \acl{WRXR} was launched on April 4, 2018 from the Kwajalein Atoll in the Marshall Islands.
Figure~\ref{fig:Rocket} shows the rocket before, during and after the launch.
The total experiment time was \SI{919}{\second} with \SI{282.5}{\second} while pointing on~target.
Both telescopes survived the~launch, observation and impact when the~rocket landed on~the~water.
A damage to the rocket fuselage was caused during descend by uneven heat dissipation.
However, despite a breach of the electronics compartment has caused a water leakage on our hermetic electronics box, the electronics inside of the box survived and data were recovered.

Analysis of the data shown that the internal pressure in the optics compartment has temporarily spiked during the launch \cite{Wages2019}.
This was presumably caused by a water vapor outgassing.
The optics and electronics section of the payload were supposed to be isolated according to the original flight plan and the optics compartment was planned to be pumped out to high vacuum.
Instead, the both sections were connected due to a last-moment failure of the ion pump for the optics section.
It is suspected that water from the payload skins caused a rise in the compartment pressure \cite{Wages2019} and could later condensed and formed ice on the instrumentation.

The~goal of~this mission was to verify the~\ac{TRL} of~the~devices which successfully worked during the~\ac{REX} mission.



\subsection{\acl{IMU}}

\begin{figure}
  \centering 
  \includegraphics[width=.95\textwidth]{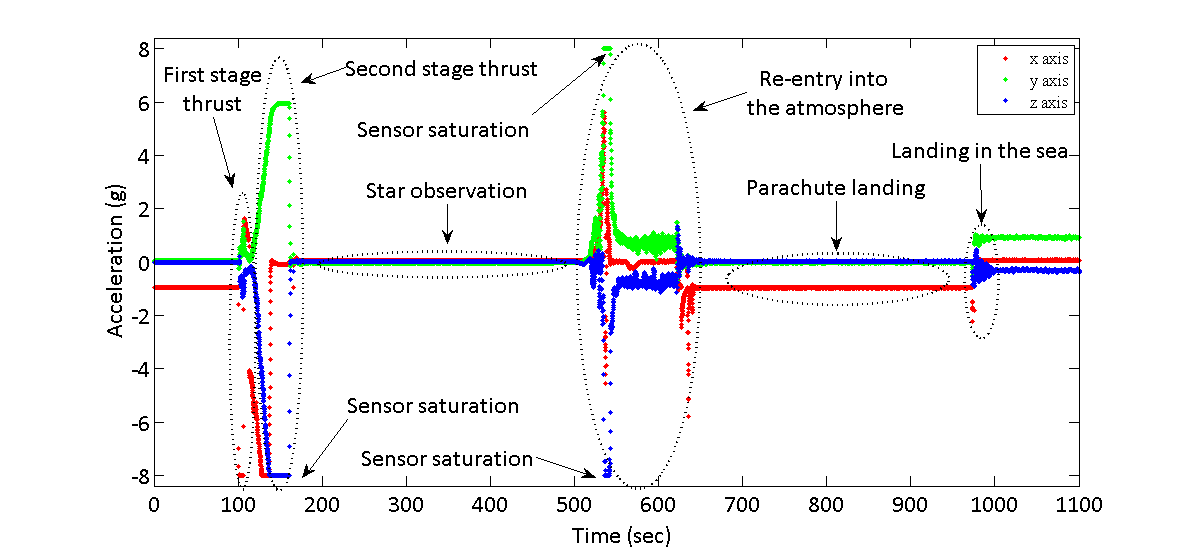}
  \caption{Data from the~onboard 3-axis \acs{MEMS} accelerometer with recognized and marked key phases of~the~suborbital mission.}
  \label{fig:IMU}
\end{figure}

The recorded data show consistent values from the~beginning of~the~mission until the~rocket parachute landed into the~ocean and stayed floating in~the~water.
All of~these facts indicate that the~sensor survived the~mission.
The~main phases of~the~flight are labeled in~\reffig{fig:IMU}, and it shows the~thrust phase of~the~first and the~second rocket engine stage, fast rocket detumbling before observation of~stars, zero gravity observing phase, reentry into the~atmosphere, parachute descending, impact onto the~water surface and free floating in~water.
The~MPU6000 sensor was partly in~saturation during the~full engine thrust due to its higher sensitivity setting, but it survived an~acceleration of~almost \SI{13.7}{\gram} and an~angular velocity of~more than \SI{3000}{deg\per\second}.
The achieved results are promising and, therefore, the MPU6000 will be used in~the~future mission of~PilsenCUBE II satellite.


\subsection{Visible camera and payload pointing}


\begin{figure*}
  \centering 
  \includegraphics[width=.95\textwidth]{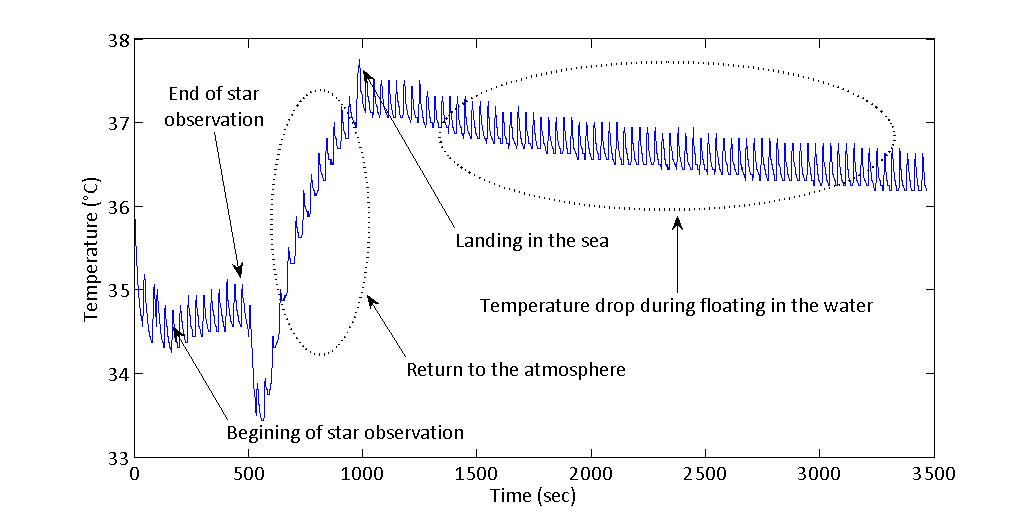}
  \caption{Logged temperature profile of~camera system during the~whole mission.}
  \label{fig:XimeaTemp}
\end{figure*}


Reliable recognition of~low brightness stars in~the~image required long exposure time and high gain of~the~image sensor due to its small pixel pitch.
However, these parameters also increase the~dark current and thermal noise which lead to the~degradation of~images, especially when the~image sensor has a~higher temperature.
Since many aspects of~the~suborbital mission were not known during the~payload preparation, we did not have simulations of~the~expected camera heating.
We proposed a~special camera capture mode with changing exposure settings in~the~loop and with idle intervals between the~individual batches of~image frames to overcome this issue of~thermal uncertainty.
The~capture mode started recording a~batch of~image frames as soon as the~rocket electronics were switched on~shortly before the~launch.
Each batch of~image frames consists of~several different combinations of~exposure time settings (from \SI{100}{\milli\second} up to \SI{900}{\milli\second}) and sensor gain settings (from \SI{0}{\decibel} up to \SI{18}{\decibel}).
The~low power mode was inserted for a~short time between two batches to reduce heating of~the~camera.
Several different exposure settings minimized the~risk of~inappropriate camera settings due to unforeseen temperature conditions and could also be used for the~combination of~images during the~post-processing.

\begin{figure}
  \centering 
  \begin{subfigure}[b]{0.32\textwidth}
    \begin{tikzpicture}
      \node[anchor=south west,inner sep=0] (a) at (0,0) { \adjincludegraphics[width=1.0\textwidth]{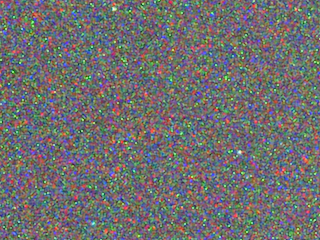}};
      \begin{scope}[x={(a.south east)},y={(a.north west)}]
        \node[imgletter] (letter) at (0.0, 0.0) {(a)};
      \end{scope}
    \end{tikzpicture}
  \end{subfigure}
  \hfill
  \begin{subfigure}[b]{0.32\textwidth}
    \begin{tikzpicture}
      \node[anchor=south west,inner sep=0] (a) at (0,0) { \adjincludegraphics[width=1.0\textwidth]{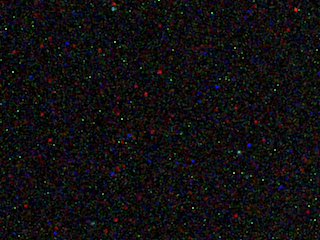}};
      \begin{scope}[x={(a.south east)},y={(a.north west)}]
        \node[imgletter] (letter) at (0.0, 0.0) {(b)};
      \end{scope}
    \end{tikzpicture}
  \end{subfigure}
  \hfill
  \begin{subfigure}[b]{0.32\textwidth}
    \begin{tikzpicture}
      \node[anchor=south west,inner sep=0] (a) at (0,0) { \adjincludegraphics[width=1.0\textwidth]{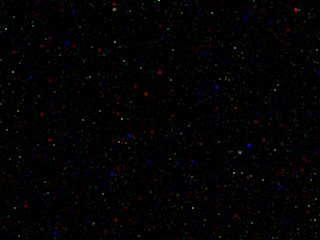}};
      \begin{scope}[x={(a.south east)},y={(a.north west)}]
        \node[imgletter] (letter) at (0.0, 0.0) {(c)};
      \end{scope}
    \end{tikzpicture}
  \end{subfigure}
  \caption{Cut-out of~the~original captured images with the~exposure time \SI{500}{\milli\second} and gains (a) \SI{18}{\decibel}, (b) \SI{6}{\decibel} and  (c) \SI{0}{\decibel}.}
  \label{fig:XimeaImages} 
\end{figure}
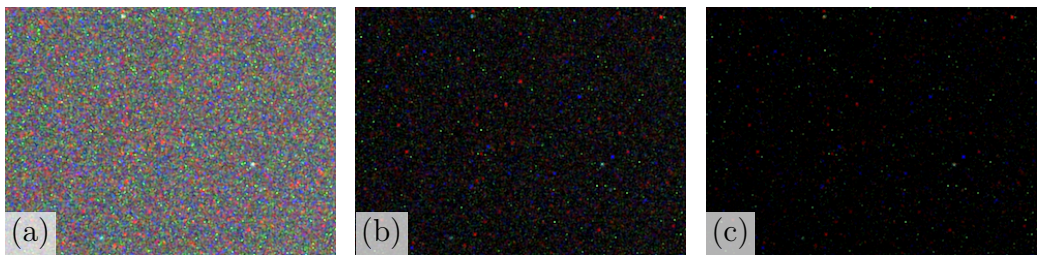

Figure~\ref{fig:XimeaImages} represents a~sample of~original images captured during the~suborbital flight, which were strongly affected by fixed pattern noise and thermal noise.
Only a~few of~the~brightest stars can be recognized, but not clearly due to strong noise artifacts in~the~images.

Camera temperature was almost constant near \SI{34.7}{\celsius} during the~observation part of~the~mission, as shown in~\reffig{fig:XimeaTemp}.
Small periodical increases and decreases of~camera temperature were caused by the~low power mode being switched on~between batches of~images to avoid overheating, but the~main temperature envelope was given by the~temperature of~mechanical structures in~the~rocket and their changes during ascending, observation, descending and the~landing phase of~the~mission.

\begin{figure}
  \centering 
  \begin{subfigure}[b]{0.32\textwidth}
    \begin{tikzpicture}
      \node[anchor=south west,inner sep=0] (a) at (0,0) { \adjincludegraphics[width=1.0\textwidth]{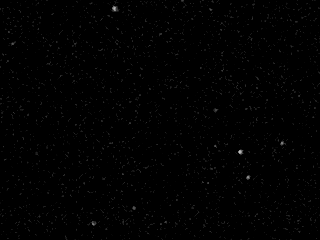}};
      \begin{scope}[x={(a.south east)},y={(a.north west)}]
        \node[imgletter] (letter) at (0.0, 0.0) {(a)};
      \end{scope}
    \end{tikzpicture}
  \end{subfigure}
  \hfill
  \begin{subfigure}[b]{0.32\textwidth}
    \begin{tikzpicture}
      \node[anchor=south west,inner sep=0] (a) at (0,0) { \adjincludegraphics[width=1.0\textwidth]{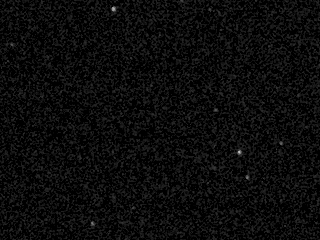}};
      \begin{scope}[x={(a.south east)},y={(a.north west)}]
        \node[imgletter] (letter) at (0.0, 0.0) {(b)};
      \end{scope}
    \end{tikzpicture}
  \end{subfigure}
  \hfill
  \begin{subfigure}[b]{0.32\textwidth}
    \begin{tikzpicture}
      \node[anchor=south west,inner sep=0] (a) at (0,0) { \adjincludegraphics[width=1.0\textwidth]{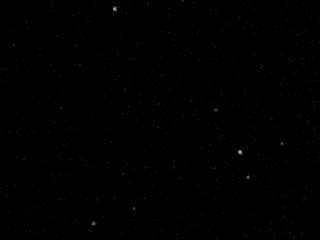}};
      \begin{scope}[x={(a.south east)},y={(a.north west)}]
        \node[imgletter] (letter) at (0.0, 0.0) {(c)};
      \end{scope}
    \end{tikzpicture}
  \end{subfigure}
  \caption{Cut-out of~the~images (same as in~\reffig{fig:XimeaImages}) after fixed pattern noise suppression by the~subtraction of~the~dark image for the~three different exposures with constant time \SI{500}{\milli\second} and gains (a) \SI{18}{\decibel}, (b) \SI{6}{\decibel} and  (c) \SI{0}{\decibel}.}
  \label{fig:XimeaImagesDark}
\end{figure}
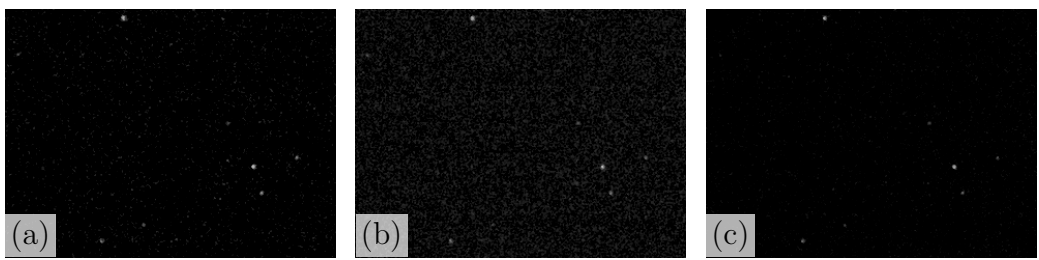

Most of~the~fixed pattern noise was removed by the~dark image subtraction, because it is almost constant for given exposure settings and camera temperature.
Dark images were captured for all exposure settings when the~rocket was in~the~ascending and descending phase of~the~mission in~a~closed instrumentation section.
Results of~dark image compensation are shown in~\reffig{fig:XimeaImagesDark}.
Nonlinear amplification and thresholding were also applied to increase the~image clarity.


\begin{figure}
  \centering 
  \begin{subfigure}[b]{0.32\textwidth}
    \begin{tikzpicture}
      \node[anchor=south west,inner sep=0] (a) at (0,0) { \adjincludegraphics[width=1.0\textwidth]{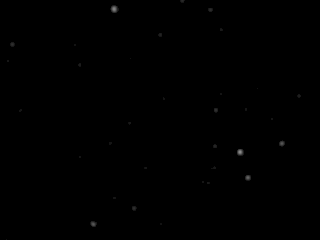}};
      \begin{scope}[x={(a.south east)},y={(a.north west)}]
        \node[imgletter] (letter) at (0.0, 0.0) {(a)};
      \end{scope}
    \end{tikzpicture}
  \end{subfigure}
  \hfill
  \begin{subfigure}[b]{0.32\textwidth}
    \begin{tikzpicture}
      \node[anchor=south west,inner sep=0] (a) at (0,0) { \adjincludegraphics[width=1.0\textwidth]{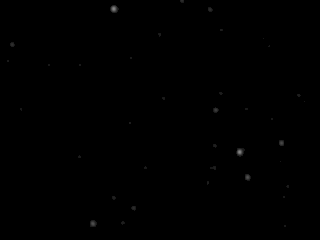}};
      \begin{scope}[x={(a.south east)},y={(a.north west)}]
        \node[imgletter] (letter) at (0.0, 0.0) {(b)};
      \end{scope}
    \end{tikzpicture}
  \end{subfigure}
  \hfill
  \begin{subfigure}[b]{0.32\textwidth}
    \begin{tikzpicture}
      \node[anchor=south west,inner sep=0] (a) at (0,0) { \adjincludegraphics[width=1.0\textwidth]{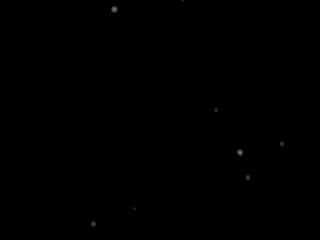}};
      \begin{scope}[x={(a.south east)},y={(a.north west)}]
        \node[imgletter] (letter) at (0.0, 0.0) {(c)};
      \end{scope}
    \end{tikzpicture}
  \end{subfigure}
  \caption{Cut-out of~the~images (same as in~\reffig{fig:XimeaImages}~and~\reffig{fig:XimeaImagesDark}) after spatial and temporal average filtering for the~three different exposures with constant time \SI{500}{\milli\second} and gains (a) \SI{18}{\decibel}, (b) \SI{6}{\decibel} and  (c) \SI{0}{\decibel}.}
  \label{fig:XimeaImagesFilter}
\end{figure}



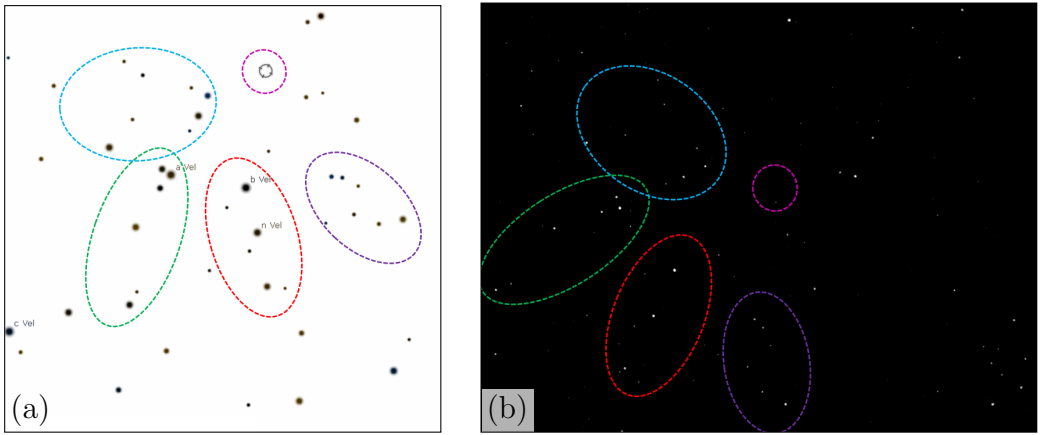
\begin{figure}
  \centering 
  \begin{subfigure}[b]{0.422\textwidth}
    \begin{tikzpicture}
      \node[anchor=south west,inner sep=0] (a) at (0,0) { \adjincludegraphics[width=1.0\textwidth]{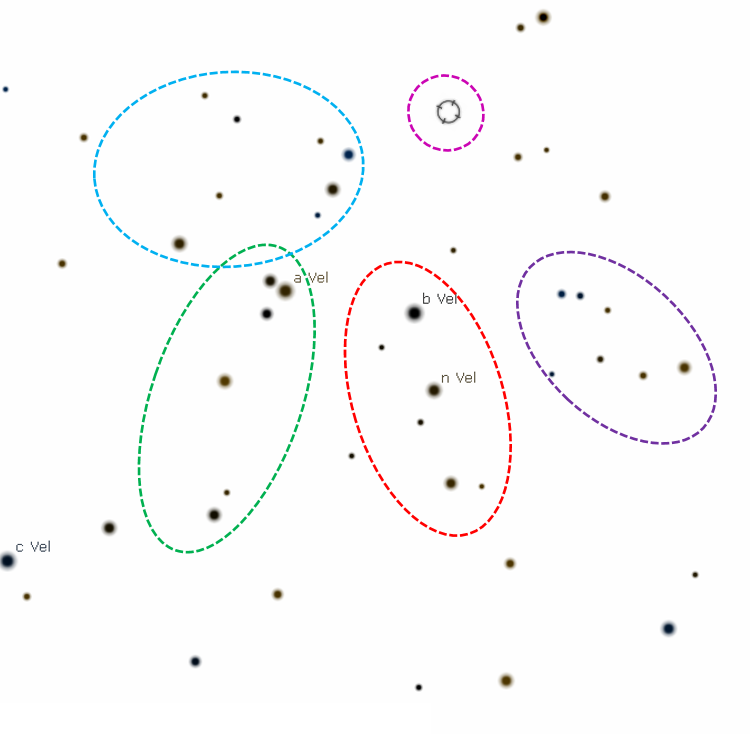}};
      \begin{scope}[x={(a.south east)},y={(a.north west)}]
        \node[imgletter] (letter) at (0.0, 0.0) {(a)};
        \draw (0.0, 0.0) rectangle (1.0, 1.0);
      \end{scope}
    \end{tikzpicture}
  \end{subfigure}
  \hfill
  \begin{subfigure}[b]{0.538\textwidth}
    \begin{tikzpicture}
      \node[anchor=south west,inner sep=0] (a) at (0,0) { \adjincludegraphics[width=1.0\textwidth]{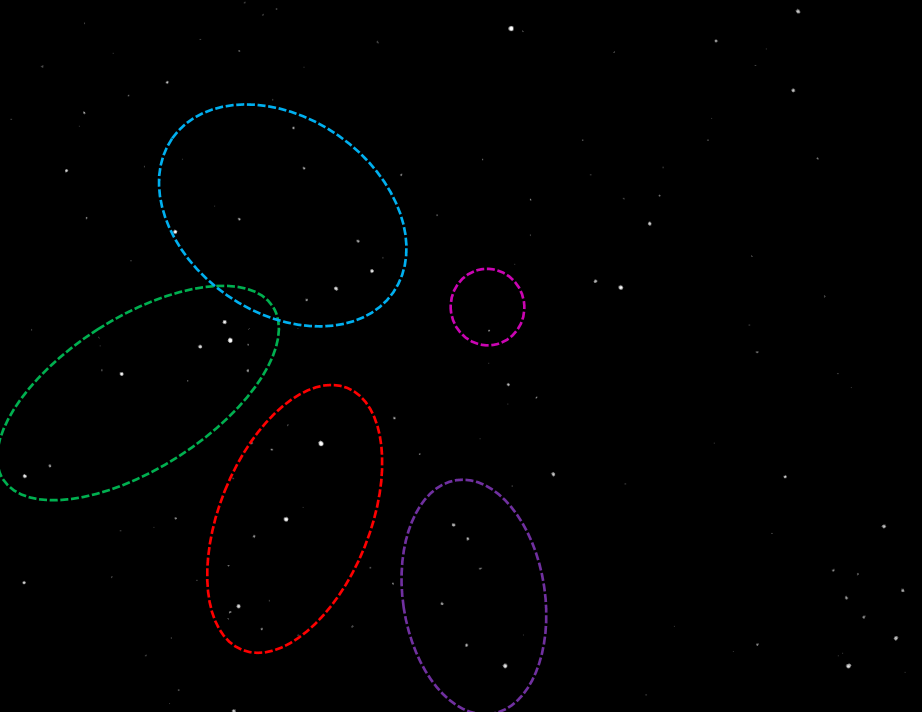}};
      \begin{scope}[x={(a.south east)},y={(a.north west)}]
        \node[imgletter] (letter) at (0.0, 0.0) {(b)};
        \draw (0.0, 0.0) rectangle (1.0, 1.0);
      \end{scope}
    \end{tikzpicture}
  \end{subfigure}
  \caption{Comparison of~star map in \emph{Stellarium} (a) and the processed images (b). The pink circle in the top-right part of (a) and in the middle of (b) was the~target of~X-ray telescopes.}
  \label{fig:XimeaRes}
\end{figure}


Bright stars can be well recognized after removing the~fixed pattern noise artifacts, however the~high amount of~thermal noise is still present in~the~images, covering and masking the~weak stars.
The~thermal noise was additionally suppressed by applying a~spatial and temporal averaging filter on~the~series of~images captured during the~attitude stabilized phase of~star observation.
The~results of~averaging eight images in~series and spatial averaging with 3\,$\times$\,3 neighboring pixels are shown in~\reffig{fig:XimeaImagesFilter}.

Most of~the~thermal noise was suppressed by averaging filters and this enabled less bright stars to be recognized (apparent magnitude 8).
Some residual fixed pattern and thermal noise is still present in~the~images, but stars can be successfully identified.
A~comparison of~the~processed image with the~\emph{Stellarium} star map shows a~good match in~\reffig{fig:XimeaRes}.

The small and cheap camera successfully fulfilled the~task in~the~suborbital mission and confirmed the~targeting of~the~X-ray telescopes.
Use of~these cameras in~small satellites like CubeSat requires capturing a~dark image and having a~long exposure time (which requires attitude stabilization).



\subsection{Infrared sensor array}

\begin{figure}
  \begin{center}
    \begin{subfigure}[b]{0.31\textwidth}
      \begin{tikzpicture}
        \node[anchor=south west,inner sep=0] (a) at (0,0) { \adjincludegraphics[width=1.0\textwidth]{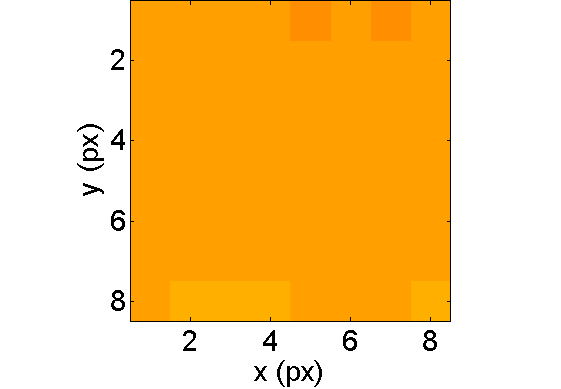}};
        \begin{scope}[x={(a.south east)},y={(a.north west)}]
          \node[imgletter] (letter) at (0.0, 0.16) {(a)};
        \end{scope}
      \end{tikzpicture}
      \label{fig:ir1}
    \end{subfigure}
    \hfill
    \begin{subfigure}[b]{0.31\textwidth}
      \begin{tikzpicture}
        \node[anchor=south west,inner sep=0] (a) at (0,0) { \adjincludegraphics[width=1.0\textwidth]{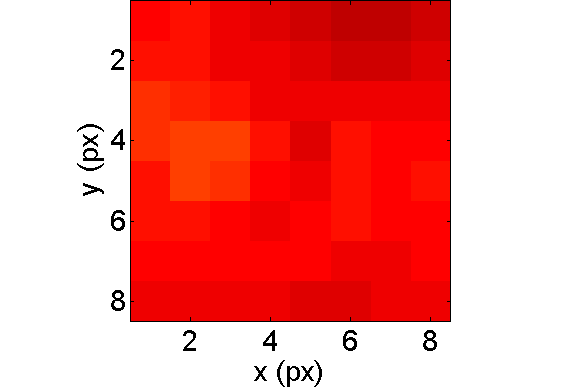}};
        \begin{scope}[x={(a.south east)},y={(a.north west)}]
          \node[imgletter] (letter) at (0.0, 0.16) {(b)};
        \end{scope}
      \end{tikzpicture}
      \label{fig:ir2}
    \end{subfigure}
    \hfill
    \begin{subfigure}[b]{0.31\textwidth}
      \begin{tikzpicture}
        \node[anchor=south west,inner sep=0] (a) at (0,0) { \adjincludegraphics[width=1.0\textwidth]{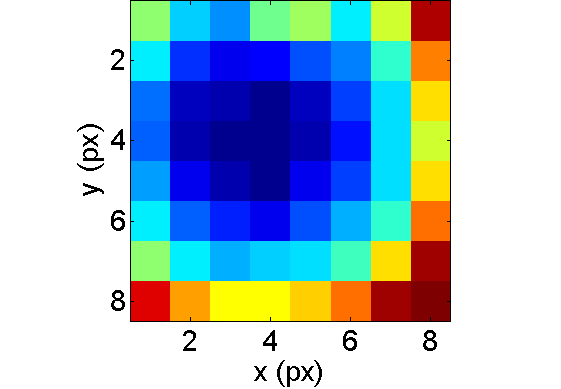}};
        \begin{scope}[x={(a.south east)},y={(a.north west)}]
          \node[imgletter] (letter) at (0.0, 0.16) {(c)};
        \end{scope}
      \end{tikzpicture}
      \label{fig:ir3}
    \end{subfigure}
    \begin{subfigure}[b]{0.31\textwidth}
      \begin{tikzpicture}
        \node[anchor=south west,inner sep=0] (a) at (0,0) { \adjincludegraphics[width=1.0\textwidth]{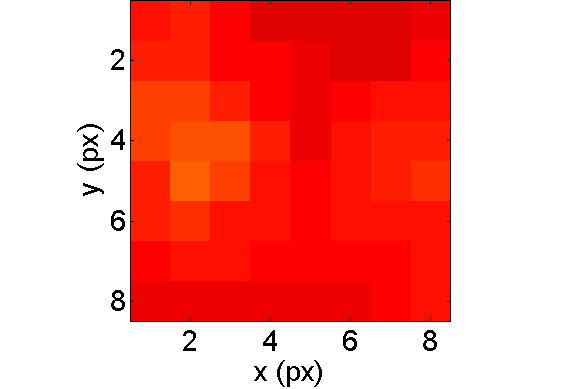}};
        \begin{scope}[x={(a.south east)},y={(a.north west)}]
          \node[imgletter] (letter) at (0.0, 0.16) {(d)};
        \end{scope}
      \end{tikzpicture}
      \label{fig:ir4}
    \end{subfigure}
    \hfill
    \begin{subfigure}[b]{0.31\textwidth}
      \begin{tikzpicture}
        \node[anchor=south west,inner sep=0] (a) at (0,0) { \adjincludegraphics[width=1.0\textwidth]{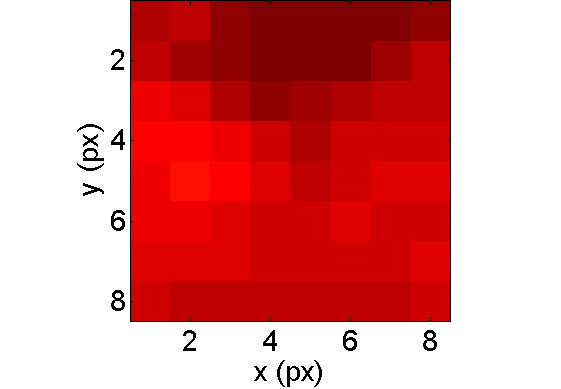}};
        \begin{scope}[x={(a.south east)},y={(a.north west)}]
          \node[imgletter] (letter) at (0.0, 0.16) {(e)};
        \end{scope}
      \end{tikzpicture}
      \label{fig:ir5}
    \end{subfigure}
    \hfill
    \begin{subfigure}[b]{0.31\textwidth}
      \begin{tikzpicture}
        \node[anchor=south west,inner sep=0] (a) at (0,0) { \adjincludegraphics[width=1.0\textwidth]{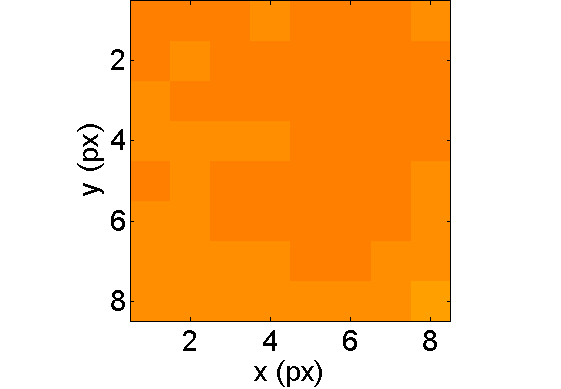}};
        \begin{scope}[x={(a.south east)},y={(a.north west)}]
          \node[imgletter] (letter) at (0.0, 0.16) {(f)};
        \end{scope}
      \end{tikzpicture}
      \label{fig:ir6}
    \end{subfigure}
    \begin{subfigure}{1.0\textwidth}
      \vspace{-1.5em}
      \small
      \begin{tikzpicture}
        \begin{axis}[
            hide axis,
            scale only axis,
            height=0pt,
            width=0pt,
            colormap/jet,
            colorbar horizontal,
            point meta min=-150,
            point meta max=100,
            colorbar style={
              xlabel=Temperature ($^\circ$C),
              width=0.85\linewidth,
              height=0.3cm,
              xtick={-150,-125,-100,...,100}
            }]
            \addplot [draw=none] coordinates {(0,0)};
        \end{axis}
      \end{tikzpicture}
      \vspace{-0.5em}
    \end{subfigure}
    \caption{Obtained thermal images from the AMG88 \ac{IR} sensor; (a) before launch, (b) during ascending, (c) after science section door opening, (d) after science section door closing, (e) during descending, and (f) after landing.}
    \label{fig:IR}
  \end{center}
  \vspace{-0.3cm}
\end{figure}
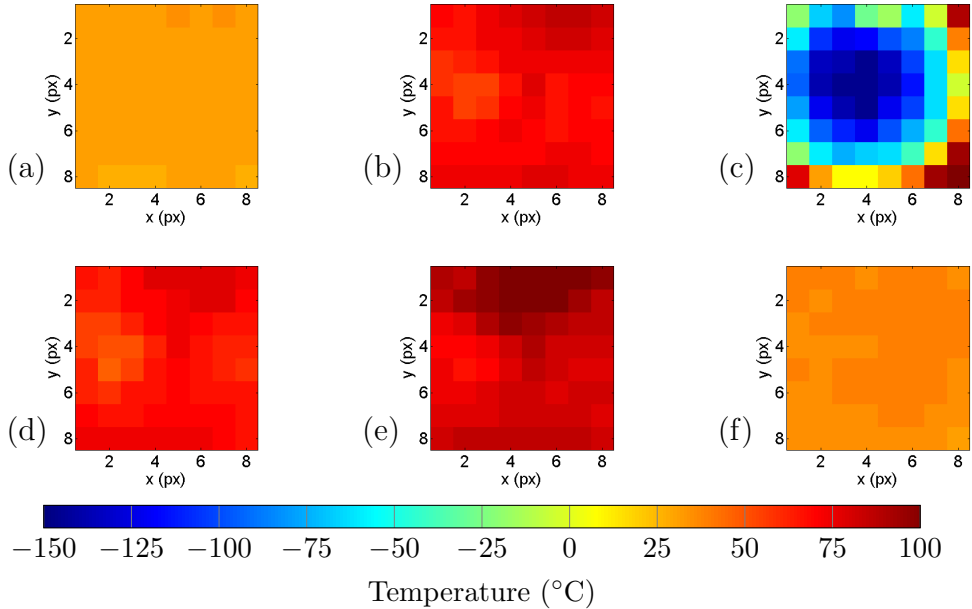

Unfortunately, the~timing of~the~rocket stabilization and door opening and closing in~the~scientific section did not show the~Earth to the \ac{FOV} of the Grid-EYE sensor, but valuable data were nevertheless obtained.
As this and similar \ac{IR} arrays contain \ac{IR} lenses, sensors should not have to experience high acceleration and the~conditions during rocket launch are far behind the~recommended operation limits.
However, the~thermal image had sharp edges during the~observation part of~the~mission so it indicates that the~\ac{IR} lenses survived the~launch conditions.
During the~ascending and descending phase of~the~mission, the~sensor's \ac{FOV} was the~door and the~mechanical structure of~the~rocket.
The~thermal data from the~sensor (\reffig{fig:IR}) indicated the~heating of~the~inner rocket structure from the~initial 25\,--\,30\,\si{\celsius} up to \SI{80}{\celsius}, due to friction of~the~atmosphere during the~ascending phase, followed by a~fast decrease of~temperature down to \SI{-150}{\celsius} (post-processed value), after the~door opening for the~phase of~star observation.
Closing the~science section door and reentry into the~atmosphere caused a~short increase of~temperature of~the~inner mechanical structure, followed by cooling during its slow parachute descent and landing into the~ocean.


\begin{figure}
  \centering 
  \begin{subfigure}[b]{0.32\textwidth}
    \includegraphics[width=\textwidth]{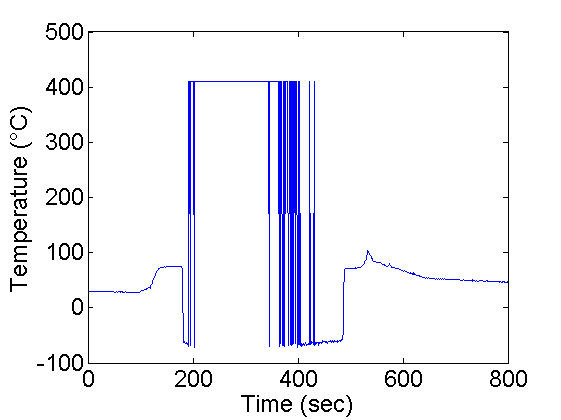}
    \caption{Pixel (2,5).}
  \end{subfigure}
  \hfill
  \begin{subfigure}[b]{0.32\textwidth}
    \includegraphics[width=\textwidth]{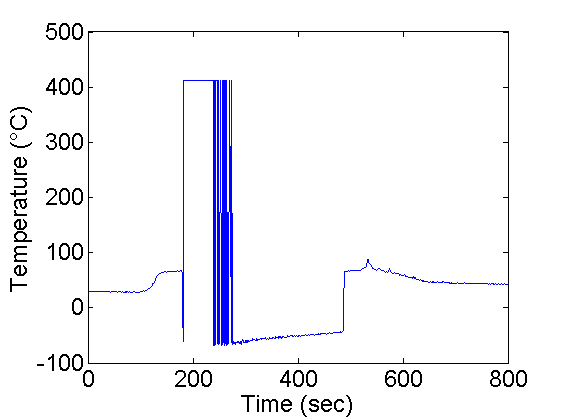}
    \caption{Pixel (6,5).}
  \end{subfigure}
  \hfill
  \begin{subfigure}[b]{0.32\textwidth}
    \includegraphics[width=\textwidth]{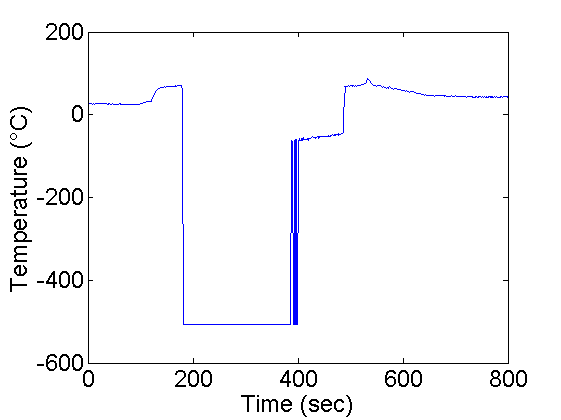}
    \caption{Pixel (8,3).}
  \end{subfigure}
  \caption{Examples of~acquired temperature measurements affected by a~software overflow in~numerical expression of~temperatures; (a) in pixel position (2,5), (b) in pixel position (6,5), (c) in pixel position (8,3).}
  \label{fig:IRtemperature}
\end{figure}


The most interesting data were expected after opening the~science section and subsequent exposure of~the~sensor to the~absence of~heat from space, and to the~heat of~the~rocket structure, both concurrently in~the~\ac{FOV} of~the~sensor, where the~sensor is exposed to a~high dynamic range of~measured temperatures.
However, the~sensor suffered from a~software overflow or sensitivity switching in~this situation and the~sensor's pixels shift numeric representation of~temperatures approximately one half of~the~numeric range, which leads to a~false interpretation of~the~measured temperatures from two complement binary representation.
This can be seen in~\reffig{fig:IRtemperature}, where the~readout temperature of~the~pixels alternates between high temperature and low temperature.

\begin{figure}
  \centering 
  \includegraphics[width=.95\textwidth]{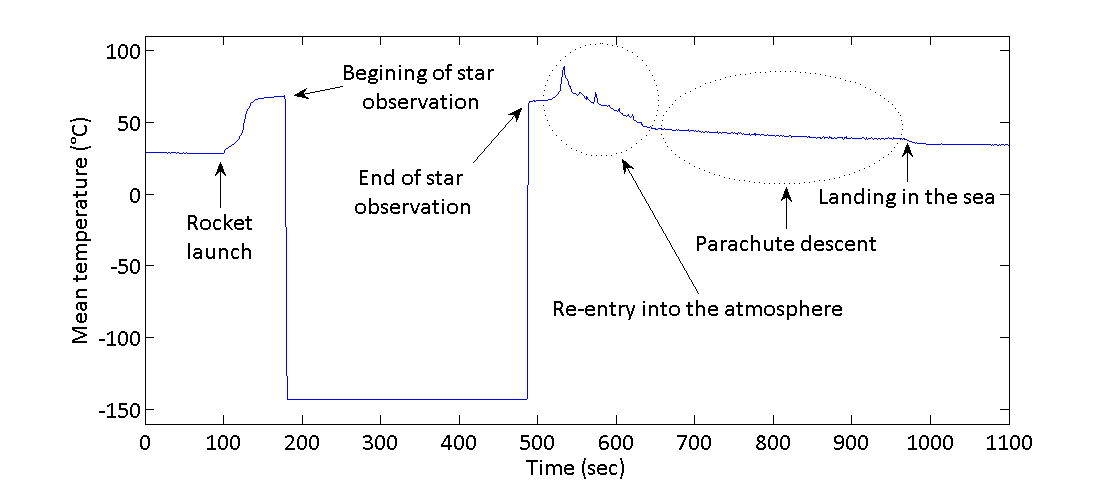}
  \caption{Temperature in~the~\acl{FOV} after compensation of~numeric representation overflow --- plotted as mean value of~four central pixels.}
  \label{fig:IRres}
\end{figure}

The overflow behavior was compensated in post-processing based on the numeric range of the temperature oscillations and knowing the expected temperature values for specific locations of the rocket structure in the sensor \ac{FOV}. Compensated temperatures with the~opened scientific section are shown in~\reffig{fig:IRres} as the~mean value of~four pixels in~the~middle of~the~AMG88 Grid-EYE sensor.


The tested \ac{IR} array sensor AMG88 is usable for simple and non-critical application in~similar missions, even with the~instability of~numeric representation of~temperatures.
The~temperature changes in~the~\ac{FOV} are detectable, only the~absolute value of~temperature is unreliable.
The~sensor can be used, for example, as a~simple detector for the~release of~the~deployable parts, covering the~sensor in~the~stowed state and opening the~sensor \ac{FOV} to space in~the~deployed state.
Using this sensor as an~Earth horizon detector in~the~\ac{ADCS} of~small satellites is not suitable with this particular sensor due to its instability of~numeric representation of~temperatures.




\subsection{X-ray telescopes}

Both X-ray detectors were autonomously operated by a custom-built software \citep{Baca2018} in \ac{ROS}.
Data from both detectors were acquired using as a~continuous stream of~\SI{1}{\second} acquisitions that started before launch.


\begin{figure}[b!]
  \centering 
  \begin{subfigure}[b]{0.48\textwidth}
    \includegraphics[width=\textwidth]{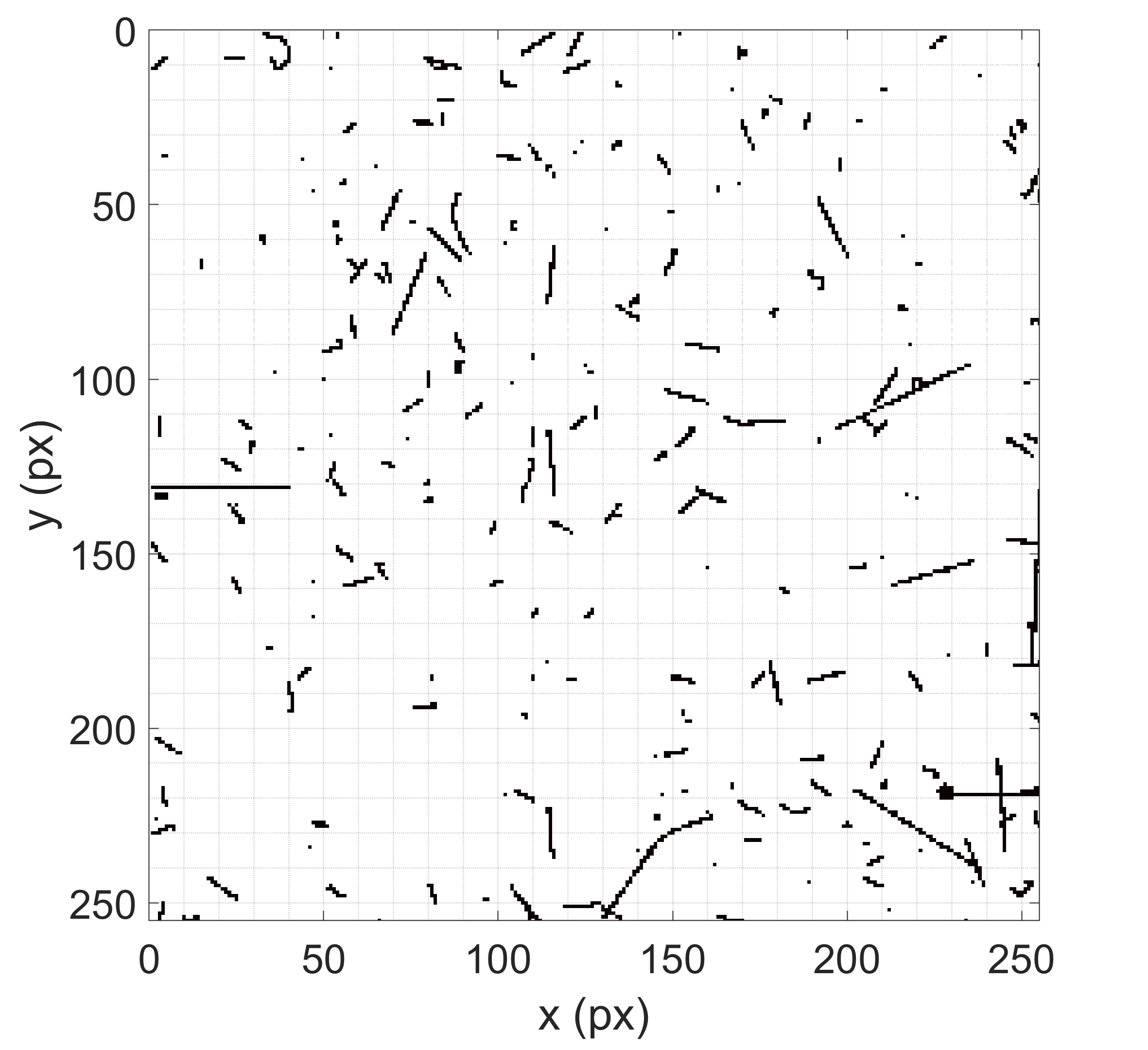}
    \caption{\acs{1D} \acs{LE} integral image.}
    \label{fig:1DExperimentSum}
  \end{subfigure}
  \hfill
  \begin{subfigure}[b]{0.48\textwidth}
    \includegraphics[width=\textwidth]{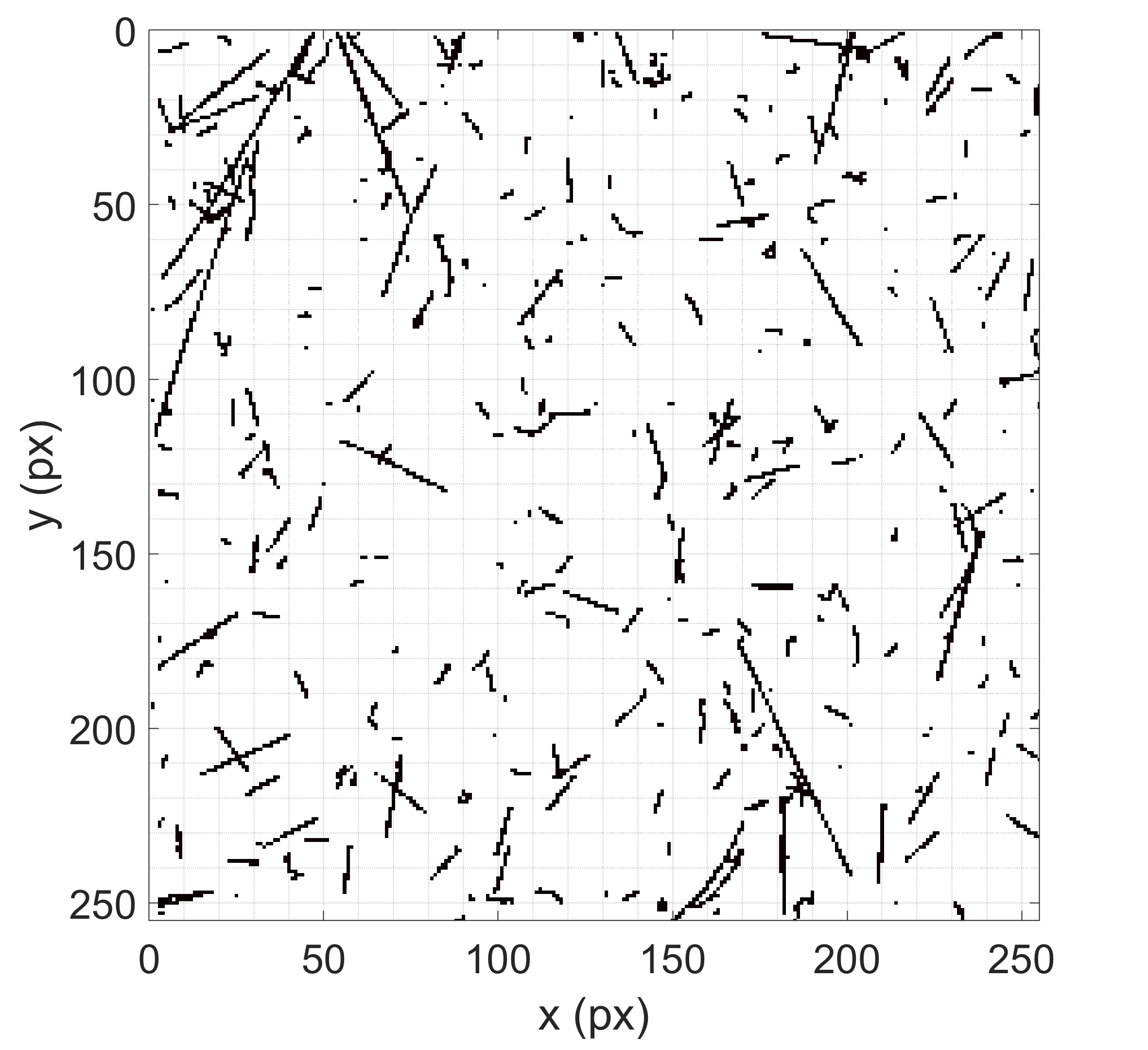}
    \caption{\acs{2D} \acs{LE} integral image.}
    \label{fig:2DExperimentSum}
  \end{subfigure}
  \caption{Integral image during the entire duration rocket's flight; (a) \acs{1D} \acs{LE} telescope, (b) \acs{2D} \acs{LE} telescope.}
  \label{fig:LEExperimentSum}
\end{figure}



\begin{figure}
  \centering 
  \begin{subfigure}[b]{0.48\textwidth}
    \includegraphics[width=\textwidth]{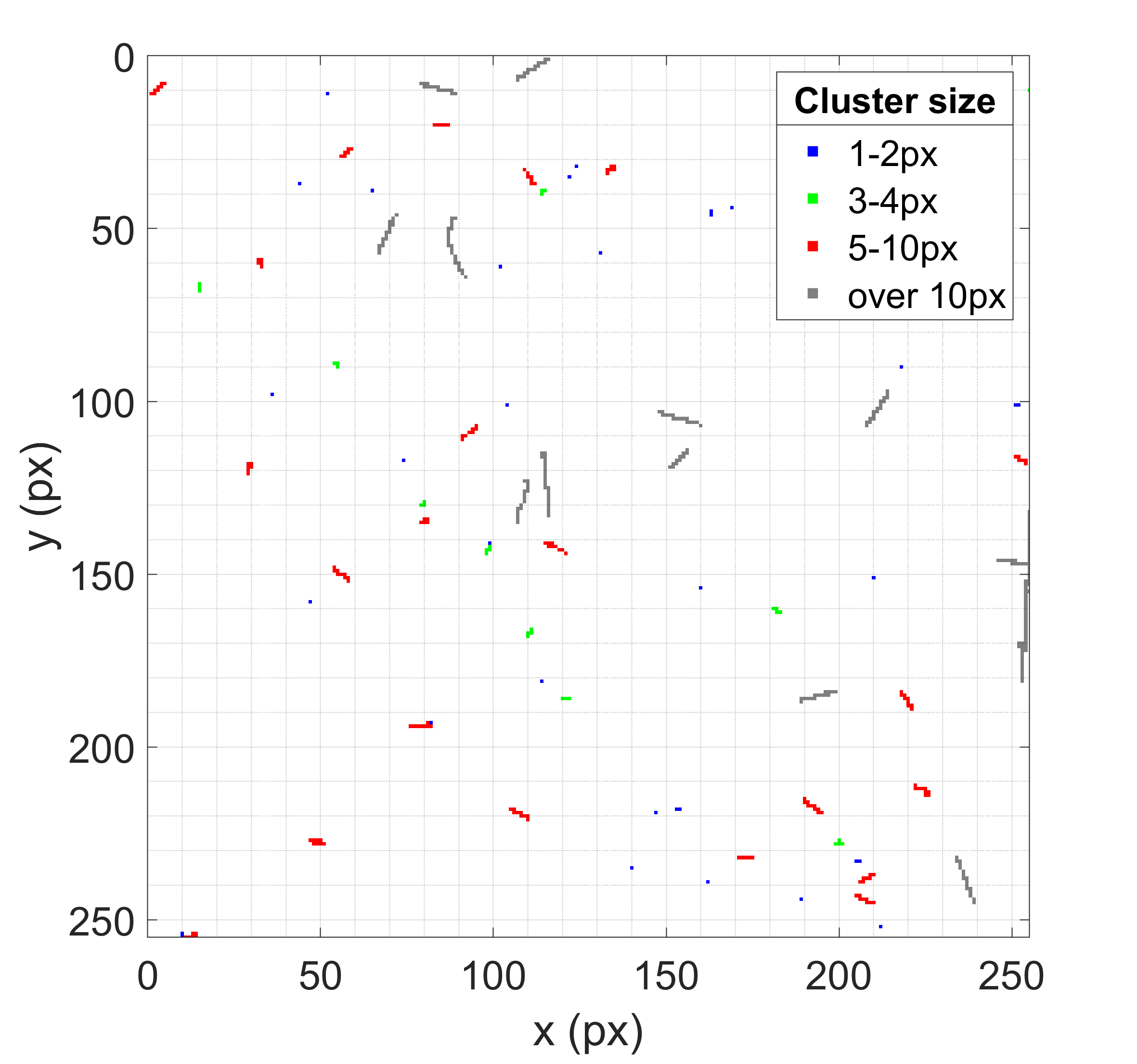}
    \caption{Integral image.}
    \label{fig:1Dimage}
  \end{subfigure}
  \hfill
  \begin{subfigure}[b]{0.48\textwidth}
    \includegraphics[width=\textwidth]{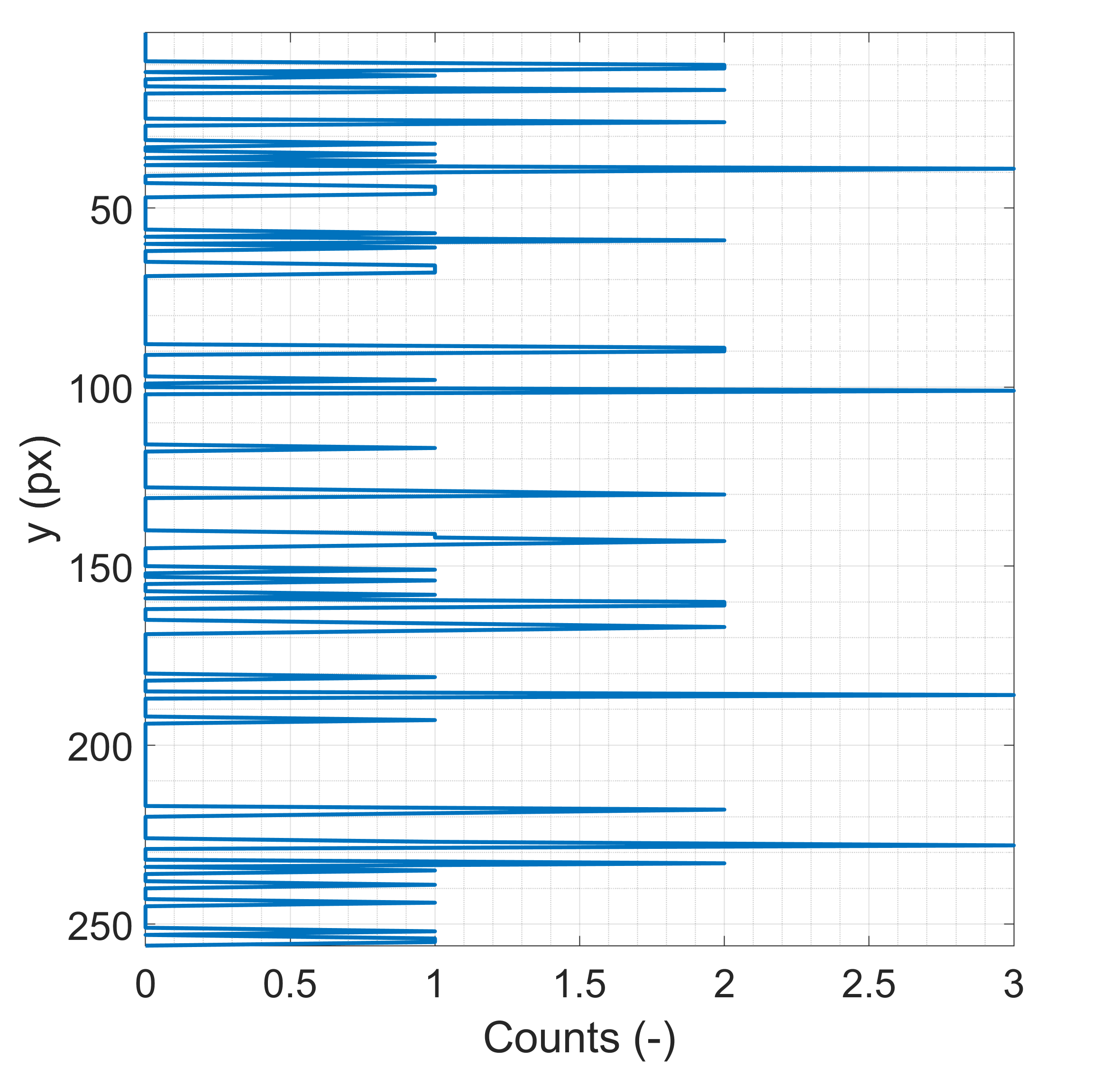}
    \caption{Row sums.}
    \label{fig:1Drows}
  \end{subfigure}
  \\
  \begin{subfigure}[b]{0.48\textwidth}
    \includegraphics[width=\textwidth]{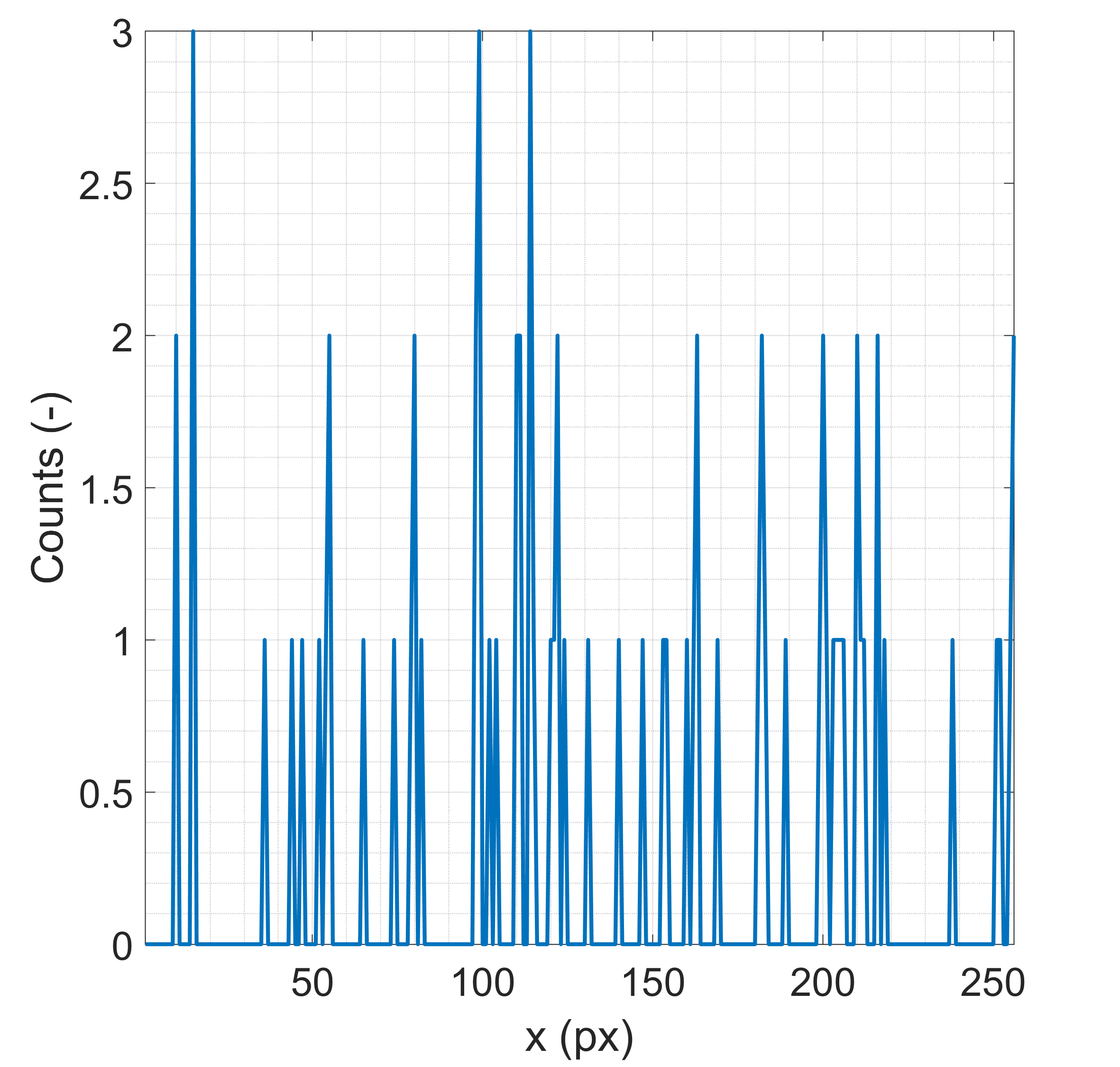}
    \caption{Column sums.}
    \label{fig:1Dcolumns}
  \end{subfigure}
  \hfill
  \begin{subfigure}[b]{0.48\textwidth}
    \includegraphics[width=\textwidth]{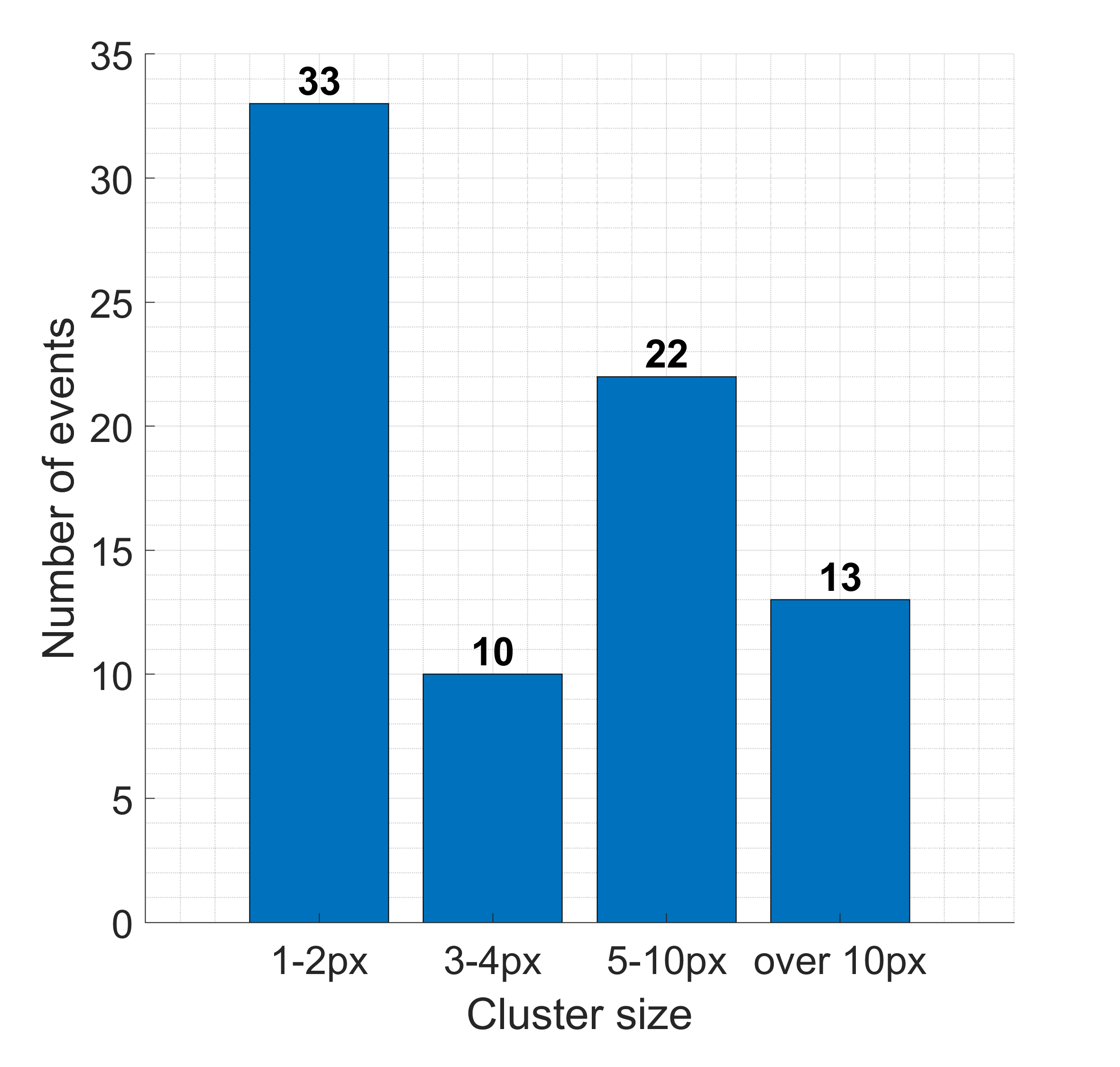}
    \caption{Cluster size histogram.}
    \label{fig:1Dbar}
  \end{subfigure}
  \caption{Data obtained from the~\acs{1D} telescope taken while pointing at~the~Vela pulsar: (a)~integral images with particle tracks labeled according to their size (b) profile of~summation of~image rows (1\,--\,4\si{\pixel}) (c) profile of~the~summation of~image columns (1\,--\,4\,\si{\pixel}) (d) bar graph representing the~frequency of~events according to their size.}
  \label{fig:1Dreal}
\end{figure}



\begin{figure}
  \centering 
  \begin{subfigure}[b]{0.48\textwidth}
    \includegraphics[width=\textwidth]{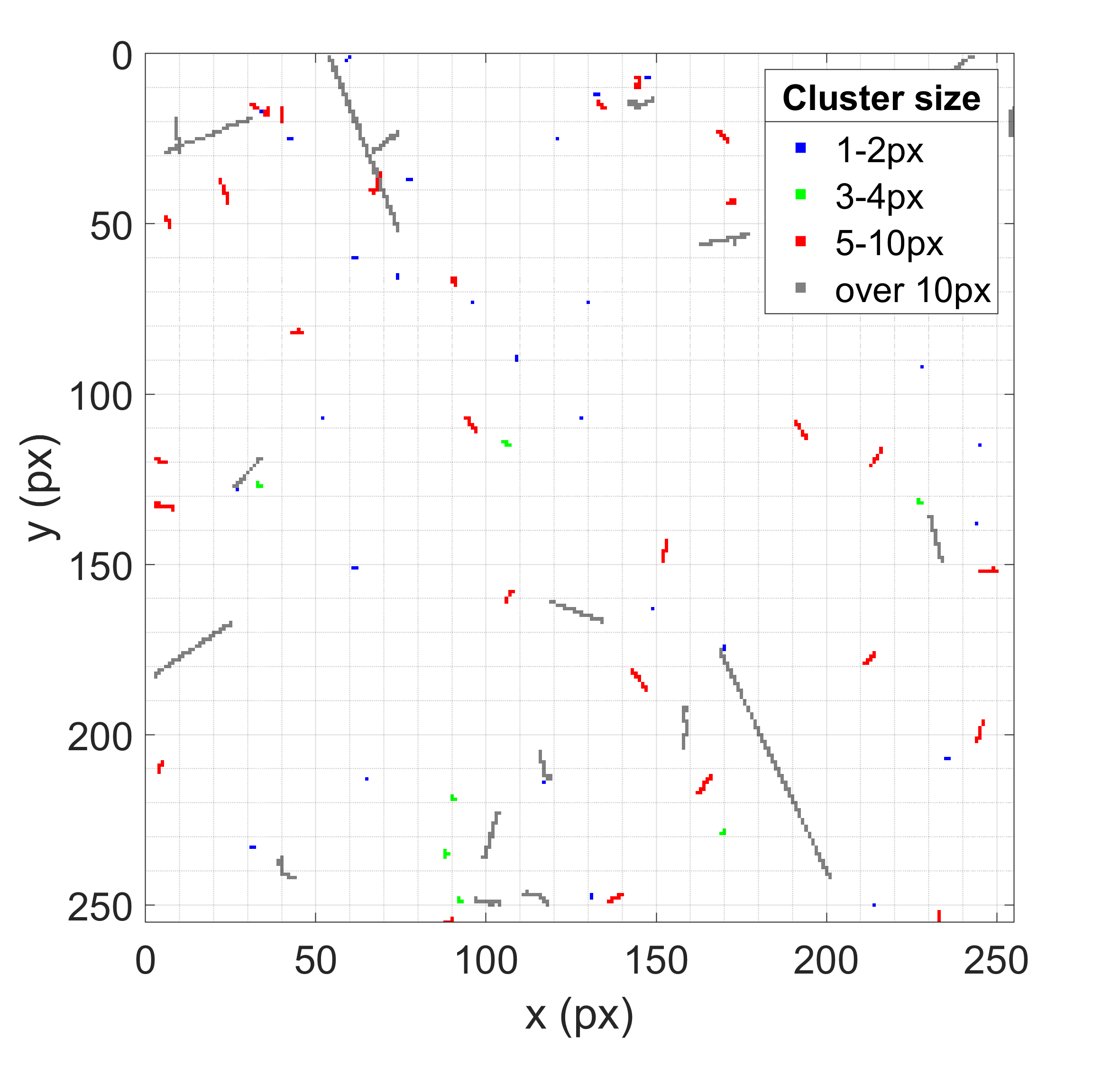}
    \caption{Integral image.}
    \label{fig:2Dimage}
  \end{subfigure}
  \hfill
  \begin{subfigure}[b]{0.48\textwidth}
    \includegraphics[width=\textwidth]{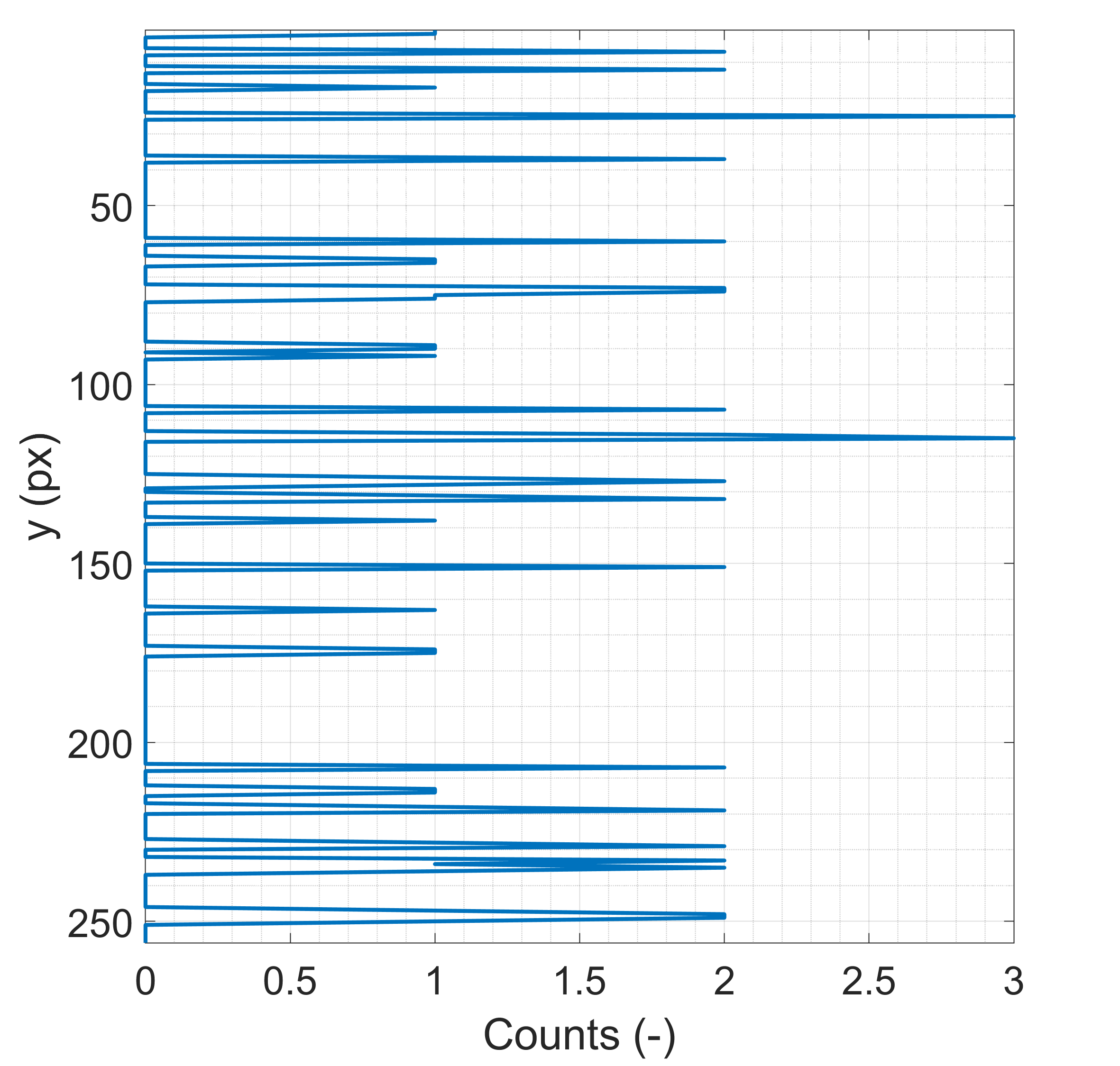}
    \caption{Row sums.}
    \label{fig:2Drows}
  \end{subfigure}
  \\
  \begin{subfigure}[b]{0.48\textwidth}
    \includegraphics[width=\textwidth]{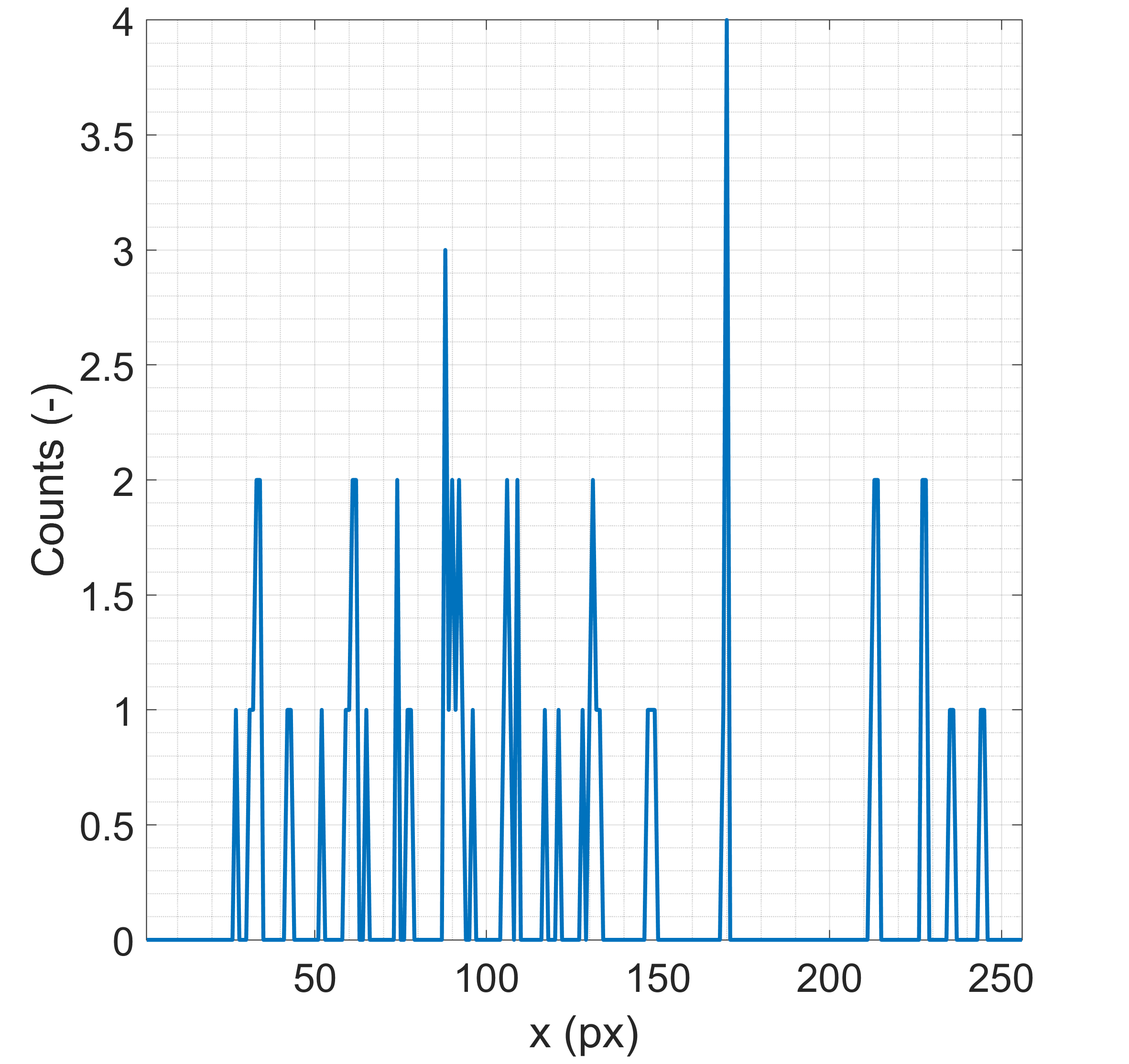}
    \caption{Column sums.}
    \label{fig:2Dcolumns}
  \end{subfigure}
  \hfill
  \begin{subfigure}[b]{0.48\textwidth}
    \includegraphics[width=\textwidth]{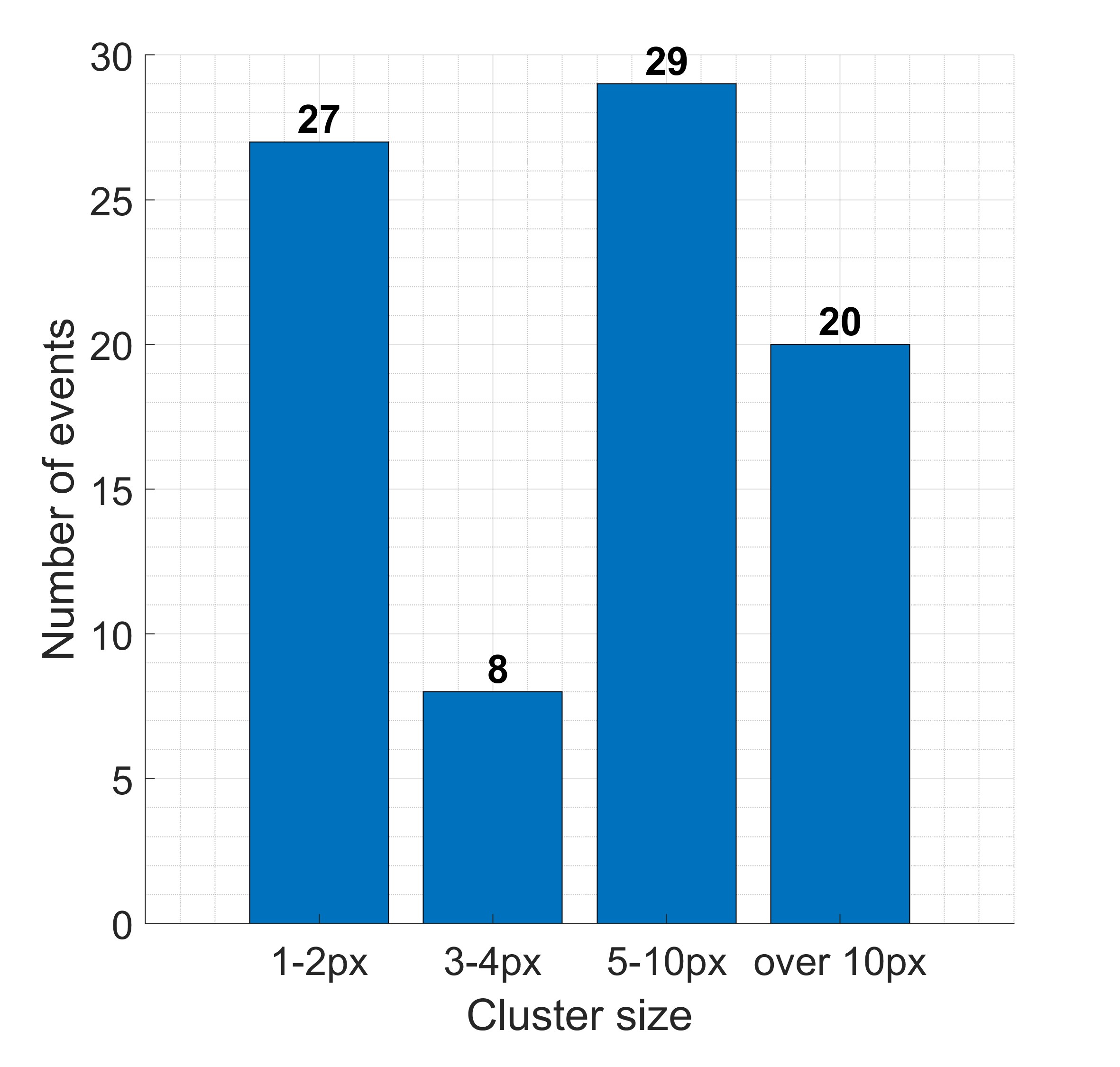}
    \caption{Cluster size histogram.}
    \label{fig:2Dbar}
  \end{subfigure}
  \caption{Data obtained from the~\acs{2D} telescope taken during the~pointing time: (a)~integral images with tracks labeled according to their size (b) profile of~row summation of~the~image (1\,--\,4\,\si{\pixel}) (c) profile of~the~column summation of~the~image (1\,--\,4\,\si{\pixel}) (d) bar graph representing the~frequency of~events according to their size.}
  \label{fig:2Dreal}
\end{figure}


Both X-ray telescopes (\ac{1D} and \ac{2D}) took images during the~entire duration of~the~rocket's flight (\reffig{fig:LEExperimentSum}).
Figures \ref{fig:1Dimage}, and \ref{fig:2Dimage} represent filtered integral images from both telescopes while the rocket pointing at~the~Vela pulsar.
Thanks to the~data sensing system, it was possible to recognize and filter out individual events based on~the~size of~their tracks or their dissipated energy in~the~detection layer.
The~results of~track size classification are shown in~color.
The~main areas of~interest were gamma particles (photons with a~range of~energies between \SI{3}{\kilo\electronvolt} to \SI{10}{\kilo\electronvolt}), which in~the~case of~the~Si detector, leave mostly single-pixel events on~the~detector \citep{Jakubek2011}.
However, photons can strike multiple pixels due to charge-sharing among the~neighboring pixels.
During the~evaluation of~the~measurement, we considered 1\,--\,4\,\si{\pixel} events as gamma photons.

The filtered images, formed only by impacts of~gamma particles, were evaluated as horizontal and vertical projections.
Sums of~individual rows and columns were plotted as a line graphs, which effectively produces integral images of 256 line-detectors.
This technique makes it easier to detect the~position the~focus line (1D) and cross (2D), since the focus is always aligned with the detector rows and columns.
Figure~\ref{fig:1Dreal} shows the processed and filter data from the \ac{1D} telescope, \reffig{fig:2Dreal} depicts the data from the \ac{2D} telescope.

As it can be seen in the images, a~small number of~particles was collected on~both detectors.
Despite the~approx. 43 (\ac{1D}) respectively 35 (\ac{2D}) X-ray gamma events taken during the~observation time, no apparent focal point/line or preferential event positions was observed.
The~total supposed X-ray counts during the~whole observation were 5~photons for the~\ac{1D}~optics and 27~photons for the~\ac{2D} optics.
The~results show no statistical tendency to form a~focus line by the~\ac{1D} telescope or focus point by the~\ac{2D} telescope. 
In fact, majority of the measured signal originates from ambient background radiation, which was recorded equally by both detectors.

Another reasons for such a small photon count yield might be a water condensation on the optics surfaces, as suggested by \cite{Wages2019}.
Water vapor condensing on the surfaces of the \ac{LE} optics foils would diminish the intensity, especially for soft X-rays.
However, the real reason is unclear as not enough data supports this hypothesis for the \ac{REX} payload.
Importantly for the \ac{TRL} evaluation, both the optics and the electronics have survived the experiment and all the subsystems were fully operational.




\section{Conclusion}\label{sec:Conclusion}

The \acl{REX} was one of~two payloads on~the~suborbital rocket campaign launched from the~Marshall Islands on April 4, 2018.
The~mission goal was to demonstrate the~space performance of~X-ray multifoil telescopes in~combination with an~uncooled pixel detector Timepix and the~verification of~the~\acl{TRL}.
The~payload contained a~number of~other sensors and devices, for observation in~the~visible, \acl{IR}, and X-ray spectra.
The~payload system was operated by the~custom-made automatic software Rospix, developed specifically for this experiment.
Due to the~low intensity of~the~source, the~measurements did not yield sufficient results for X-ray observations of~the~Vela \acl{SNR}, however, we can report that all instruments in~the~proposed system worked successfully and without problems during the~experiment.
Based on~the~obtained results, the~functionality and the~possibility of~using the~proposed system for similar upcoming experiments was confirmed.



\section*{Acknowledgements}

The team of~authors of~this article would like to thank the~whole team of~colleagues who participated in~this project, not only: Ondřej Petr (\acl{RITE} s.r.o.), Robert Filgas (\acl{CTU}), Richard Linhart (\acl{UWB}) and Pavel Fiala (\acl{UWB}) and many others without which this project did not arise.

This project was performed in~collaboration with several universities, companies and scientific institutions in~the~Czech Republic and abroad.
Namely: Czech Technical University in~Prague (CZE) - Institute of Experimental and Applied Physics and Faculty of Electrical Engineering,  University of~West Bohemia (CZE), \acl{RITE} s.r.o. (CZE), Czech Aerospace Research Centre a.s. (CZE), PANTER facility of~the~Max Planck Institute for Extraterrestrial Physics (DEU), Pennsylvania State University (USA), University of~Colorado at~Boulder (USA).

Many thanks for the~willingness, help and pieces of~advice belonging to namely: D. Burrows with his team (\ac{PSU}), W. Cash with teams (CU Boulder), V. Burwitz with the~team (PANTER).

The project was supported by the~Technology Agency of~the~Czech Republic under grants no: TA03011329, TA04011295, the Grant Agency of the Czech Republic under grant no. 18-10088Y,  the NASA Astrophysics Research and Analysis grant NNX17AC88G, the~Grant Agency of the Czech Technical University in Prague no. SGS18/186/OHK3/3T/13, and by the Ministry of Education and Youth of the Czech Republic, project \linebreak no. LTAUSA18094. This work was done on behalf of Medipix2 collaboration.


\bibliographystyle{elsarticle-num-names}
\bibliography{bibliography}

\end{document}